\documentclass[10pt]{article}
\pdfoutput=1
\usepackage[left=2.54cm,top=2.54cm,right=2.54cm,bottom=2.54cm]{geometry}
\usepackage{fancyhdr}
\usepackage{afterpage}
\usepackage{setspace}
\usepackage{bibspacing}
\usepackage{amsmath}
\usepackage{float}
\newfont{\toto}{msbm10 at 12 pt}
\newfont{\ithd}{cmr9}

\usepackage{amsmath,amsthm,amsfonts,amssymb}
\usepackage[pdftex]{graphicx}
\usepackage[T1]{fontenc}

\title{
\bf Higher-Order Multilevel Framework for ADER Scheme in Computational Aeroacoustics }
\author{
	S.~M.~Joshi\footnote{email: smjoshi@aero.iitb.ac.in} ~\& A.~Chatterjee\\\\
Department of Aerospace Engineering\\
Indian Institute of Technology, Bombay\\
Mumbai 400076, India}

\date{}

\begin{document}

\maketitle
\afterpage{\fancyhead{}}

\centerline{
\begin{minipage}[t]{150mm}
{\bf Abstract:} 
The versatile Arbitrary-DERivative (ADER) scheme is cast in a multilevel framework (ML-ADER) for 
fast solution of system of linear hyperbolic partial differential equations.
The solution is cycled through spatial operators of varying 
accuracy while retaining highest-order accuracy by the use of a forcing function. 
Accuracy analysis of the multilevel framework including in the ML-ADER form is carried out in
time-domain as well as frequency-domain. 
Results are obtained for benchmark problems in computational aeroacoustics at a much reduced 
computational cost.
\end{minipage}
}
\vskip0.5cm

\section{Introduction}

Engineering problems in acoustics or electromagnetics
are often characterized by large domain size, long simulation time, large spectral bandwidth 
and complex geometrical boundaries \cite{Tam2012}.
Computational Aeroacoustics (CAA)
and Computational Electromagnetics (CEM) deal with solving such problems using numerical
methods in discrete space and time.
Majority of applications in CAA or CEM involve traveling linear waves governed by
linear hyperbolic PDEs. 
Accurate solution of these PDEs require numerical schemes with very low dispersion (phase error) 
and dissipation errors (amplitude error). 
It is seen that, spatially higher-order accurate numerical schemes reduce diffusion and dispersion errors 
over an increased range of wavenumbers \cite{Tam2012}. 
Numerical schemes can also be optimized for minimum dispersion error
over a prescribed range of frequencies. 
Such optimized schemes usually work well for signals with significant large-wavenumbers
contribution. However, benefits of optimization may reduce if the wave is well resolved or if 
the signal mainly consists of lower wavenumbers \cite{Zingg2000}.
In contrast to this, uniformly spatially higher-order accurate 
schemes are seen to perform better in case of well-resolved waves \cite{Zingg2000, Cunha2014}.
Uniformly spatially higher-order accurate schemes also 
show a more relaxed Points Per-Wavelength (PPW) requirement
as compared to lower-order (first-order or second-order) schemes.
In addition to this, PPW requirement of a higher-order scheme for satisfactory 
resolution of waves remains mostly 
unchanged even in case of a large domain size, 
making it possible to use a relatively coarser grid for achieving required resolution 
\cite{Zingg2000}.
However, higher-order schemes often use a wider stencil or more number of data-points for
numerical approximation of derivatives resulting in higher cost per-grid point. 
Moreover, the computing cost can rise significantly with increase in complexity of the problem. 
In this work we develop an algorithm for fast computation of traveling linear waves  
with higher-order accuracy at a lower computing cost.

Multi-Grid (MG) methods are commonly used for accelerating convergence to steady state
solution of boundary value problems using iterative solvers \cite{Brandt1977}. 
This is achieved by rapid elimination of high wavenumber components of the error by 
successive projection of the solution with hierarchy of approximations. The error can be reduced 
by cycling the solution through a hierarchy of successively coarser grids ($h$-MG) or 
through use of successively lower-order reconstruction operators ($p$-MG) or through
a combination of both. However, MG methods based on smoothing properties of iterative solvers
are in principle better suited for accelerating convergence to steady state and therefore may not be 
appropriate for time-accurate solution of traveling waves governed by linear hyperbolic PDEs. 
For economically computing higher-order accurate solution to problems involving traveling linear waves,
a multilevel (ML) framework was proposed in Ref.\cite{Chatterjee2015}.
In this framework, 
higher-order accuracy is inexpensively maintained at coarser approximations for an advection
process characterizing linear wave propagation involving sufficiently smooth solution.
This is achieved by cycling the solution
through successively lower-order reconstruction operators on a fixed grid resembling a $p$-MG method.
The relative truncation error $\tau$ between the higher-order and the 
lower-order accurate approximations is added to the lower-order accurate approximation 
to retain higher-order accuracy. 
The multilevel framework using Essentially-Non-Oscillatory
(ENO) reconstruction was successfully used for solving problems in CEM and CAA 
\cite{Chatterjee2015,Joshi2016b,Chatterjee2014}.
The ENO family of numerical schemes allows an unified and easy access to successive higher-order 
accuracy which is required for the ML framework. 
However, ENO schemes are ideally suited for shock dominated flows and may not be
appropriate for computations involving linear waves. 
Schemes using sub-cell resolution such as the Spectral Volume Method (SVM) work well for 
linear waves, however,  
cycling through orders of accuracy or $p$-levels requires restriction and prolongation
processes which smoothens the data and fails to retain higher-order accuracy \cite{Joshi2015}.

In this work, we develop the ML method for the Arbitrary-higher-order-DERivatives (ADER) scheme.
In the ADER scheme, spatial accuracy and temporal accuracy are coupled through the 
Cauchy-Kowalewsky procedure \cite{Titarev2002,Toro2006}. 
The ADER scheme is found to be suitable for solving linear hyperbolic systems due to superior
dispersion characteristics
\cite{Schwartzkopff2002,Schwartzkopff2004}.
A preliminary study of the multilevel ADER (ML-ADER) method was presented in Ref.\cite{Joshi2016b}. 
In this work we expand this study to include an extensive analysis of the general ML framework 
including ML-ADER in time and frequency domain.  
We also present analysis and results for a variant of ML-ADER using multiple frozen steps at the 
lowest-order approximation (frozen $\tau$). In this variant, multiple steps ($\nu > 1$) are performed 
at the lowest-order approximation while maintaining higher-order accuracy by addition of a 
frozen truncation error at each time step, thereby further reducing computing time. 
The more common sawtooth 
cycle in MG methods turns out to be a special case with $\nu=1$. 
Finally, results for quasi $1$D and $2$D problems in CAA are presented along with cost analysis
of the method.

\section{ML-ADER Method}
\subsection{Multilevel Method}\label{ml}
In this section, we briefly discuss the ML method presented in Ref.\cite{Chatterjee2015}.
Application of the ML method to the ADER scheme is discussed in the subsequent sections.
Consider a system of hyperbolic Partial Differential Equations (PDEs) in $1$D,
\begin{equation}\label{hyper}
	{\bf U}_t +  {\bf F}_x = 0
\end{equation}
where, ${\bf U} = [u_1, u_2, ... , u_n]^T $ is a vector of conserved variables and
${\bf F} =[f_1({\bf U}), f_2({\bf U}), f_3({\bf U}), ..., f_n({\bf U})]^T$ 
is the flux vector.
In order to solve the above system of hyperbolic PDEs, the $1$D domain $\Omega$ is discretized into
$N$ non-overlapping finite volumes (FV) 
$$\Omega = \bigcup_{i=0}^{N-1}(x_{i-\frac{1}{2}},x_{i+\frac{1}{2}})$$
The volume of $i^{th}$ cell is  $\Delta x_i = \left(x_{i+\frac{1}{2}} - x_{i-\frac{1}{2}}\right)$ and 
the location of the cell-center is  $ (x_{i+\frac{1}{2}} + x_{i-\frac{1}{2}})/{2}$.
The semi-discrete formulation of equation\eqref{hyper} resulting from the finite 
volume approximation in the $i^{th}$ cell can be written as,

\begin{equation}\label{eq:ode}
	\frac{d{\bf U}_i}{dt} = \mathcal{R}^{(n)}_{i,p} =  \frac{-1}{\Delta x_i} ({\mathcal F}_{i+\frac{1}{2},p}^{(n)}- {\mathcal F}_{i-\frac{1}{2},p}^{(n)})  
\end{equation}
where ${\mathcal F_{i+\frac{1}{2},p}^{(n)}}$ is the $p^{th}$ order numerical flux at $\left(i+\frac{1}{2}\right)^{th}$ face
at time level $n$.  
A $(p-1)^{th}$ order interpolation-polynomial is used for data reconstruction resulting in
a $p^{th}$ order accurate scheme.
For $p=1$, equation\eqref{eq:ode} reduces to the traditional first-order finite volume scheme. 
Thus, $\mathcal{R}^{(n)}_{i,p}$ is the $p^{th}$-order flux-residual in the $i^{th}$ finite volume 
at time level $n$. 
It is to be noted that, $(.)^n$ indicates $n^{th}$ exponent whereas, $(.)^{(n)}$ 
indicates time-level $n$.
In the conventional finite volume numerical schemes, equation({\ref{eq:ode}}) is integrated 
numerically to advance the solution in time. At each time-step, a $(p-1)^{th}$ order polynomial is 
used for data reconstruction. This results in an uniformly $p^{th}$-order accurate 
FV scheme. 

In the ML method, the solution is cycled through a hierarchy of spatial operators of varying
order of accuracy on a constant grid similar to $p$-MG cycles.
Starting from the highest spatial order of accuracy $m$, 
each consecutive iteration is performed at subsequent lower order spatial approximation till the 
lowest spatial order of accuracy is reached.
Thus, after computing $R_{i,m}^{(0)}$ at the starting iteration,
for the next iteration $R_{i,m-1}^{(1)}$ is computed instead of $R_{i,m}^{(1)}$.  
The relative truncation error between the $m^{th}$ order approximation and a 
local lower-order approximation is added to the lower-order flux residual
to restore the highest-order accuracy.
The relative truncation error ($\tau$) between the highest ($m^{th}$) and the local
$p^{th}$-order approximations at time level $n$ is defined as
\begin{equation}\label{tau}
	\tau_{i,p}^{(n)} = \mathcal{R}_{i,p+1}^{(n-1)} - \mathcal{R}_{i,p}^{(n-1)} + \tau_{i,p+1}^{(n-1)}
\end{equation}
It is to be noted that, at the highest level ($p=m$), the forcing function $\tau_{i,m}^{(.)}$ is a 
zero vector.
This relative truncation error $\tau_{i,p}^{(n)}$ is added  to $\mathcal{R}^{(n)}_{i,p}$ to get a 
modified residual $\mathcal{R}_{i,mod}^{(n)}$.
This modified residual at time level $n$ is thus written as, 
$$\mathcal{R}_{i,mod}^{(n)} = \mathcal{R}_{i,p}^{(n)} + \tau_{i,p}^{(n)}$$
and the resultant ODE in $i^{th}$ FV is then written as,
\begin{equation}\label{eq:ode1}
	\frac{d{\bf U}_i}{dt}= \mathcal{R}_{i,mod}^{(n)}
\end{equation}

\begin{figure}[h]
	\vspace{4mm}
	\begin{center}
	\includegraphics [scale=0.5]{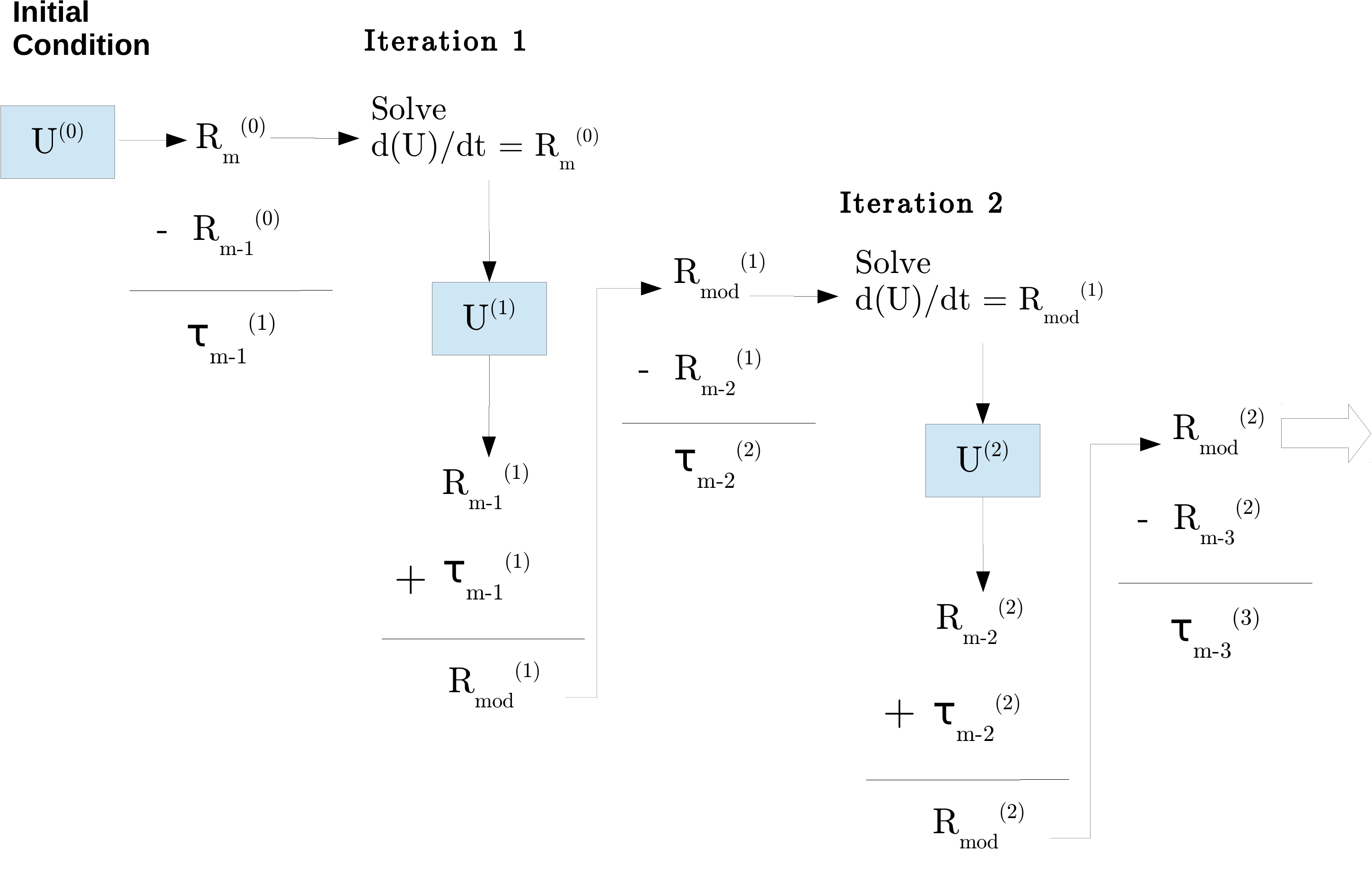}
	\end{center}
	\caption{\label{MLbetter} ML Algorithm: Addition of the Truncation Error as the Forcing Function }
\end{figure}
\begin{figure}[h!]
	\begin{center}
		\includegraphics [scale=0.55]{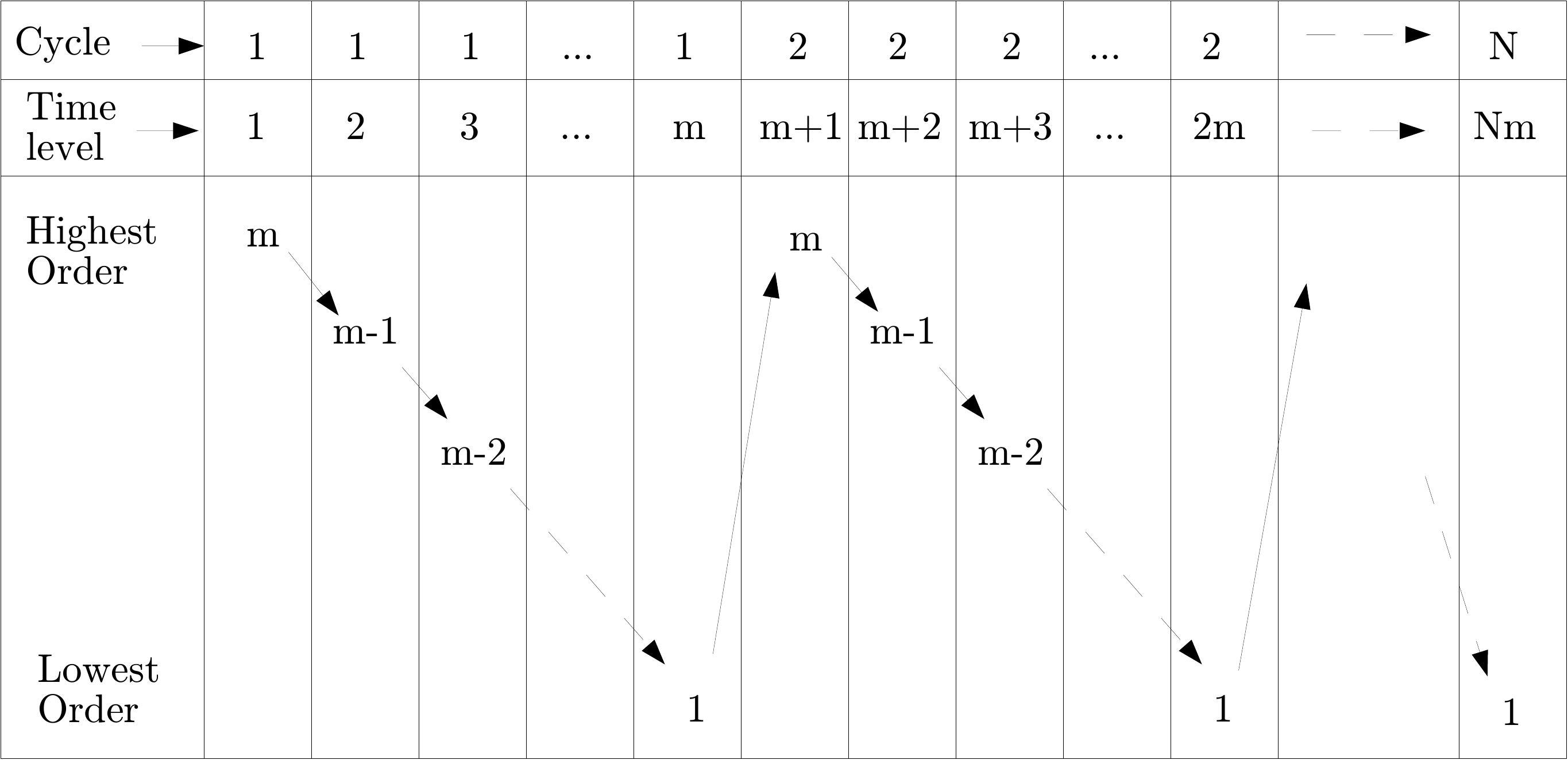}
	\end{center}
	\caption{\label{mlbetter} ML Algorithm: Cycling the Solution Through Successively Coarser Approximations}
\end{figure}

Equation(\ref{eq:ode1}) is integrated numerically to advance the solution to time level $n+1$.
This process is illustrated in Fig.\ref{MLbetter}.
At the lowest-order spatial accuracy, an appropriate strategy can be used to return to the 
highest-level approximation. Fig.\ref{mlbetter} shows the `sawtooth' cycle used in the ML method.

As shown subsequently, the resulting solution turns out to be spatially 
$m^{th}$-order accurate if the data is `sufficiently smooth' \cite{Chatterjee2015}.
`Sufficiently smooth' translates into condition of local linearity for a two-level second-order method
\cite{Chatterjee2015}.
Moreover, it can be argued that at all levels of approximation, highest-order accuracy
can be retained by successive addition of $\tau$ as a forcing function to lower-order residuals.  
Analytical proof for a two-level ADER method is presented in the following section.

\subsection{ML-ADER Implementation}

The ADER scheme ({\bf A}rbitrary higher-order DERivatives)
 is based on the MGRP (Modified Generalized Riemann Problem) 
scheme of Toro \cite{Toro1998}; which in turn is a simplification of the Generalized Riemann Problem (GRP) scheme of
Ben-Artzi and Falcovitz \cite{Ben-Artzi1984}. 
Schwartzkopff and Toro developed the ADER scheme for 
linear systems \cite{Schwartzkopff2002,Schwartzkopff2004}.
Application of the ADER scheme to non-linear hyperbolic problems, as well as 3D generalization 
were carried out 
by Titarev and Toro \cite{Titarev2002,Toro2005, Titarev2005}.
ADER schemes are found to be especially suitable for wave propagation because of dampening of the 
dispersive modes 
thereby preserving wave phase and 
amplitude \cite{Schwartzkopff2002}. A brief description of the ADER method is presented below.

A Riemann problem arising from piecewise polynomial data is known as the Derivative Riemann
Problem (DRP) \cite{Toro2006}. 
Solution of the DRP yields the state variable, as well as higher-order spatial derivatives of the 
state variable at cell interfaces. Corresponding time-derivatives can be obtained
using the Lax-Wendroff procedure which links time-derivatives to the spatial derivatives.
Time-derivatives of the state variable at interfaces in turn can be used for
the Taylor series approximation of time-evolution of the state variable. 
Thus, in the ADER scheme, higher-order approximation to the state variable in increasing time $t$
is obtained by combining higher-order spatial reconstruction and finding solution of the DRP at cell
interfaces\cite{Toro2006}. 
The ($K^{th}$-order) Taylor series expansion of the state variable ${\bf U}(x,t)$ at face $i+\frac{1}{2}$ is written as,

\begin{equation}\label{UR}
	{\bf \tilde U}(x_{i + \frac{1}{2}},\tau) = \underbrace{{\bf U}(x_{i + \frac{1}{2}},0_+)}_{1} + \underbrace{\sum_{k=1}^{K} \left[\partial_{t}^{k}{ {\bf U}(x_{i + \frac{1}{2}},0_+)}\right] 
	\frac{\tau^{k}}{k!} }_{2}
\end{equation}

where, the first term on the right hand side is the 
solution of the conventional Riemann problem based on the discontinuity in the
state variable at cell interfaces (Godunov state). The second term is obtained by first 
converting the time derivatives into spatial derivatives using the Lax-Wendroff procedure and
then solving $K-1$ DRPs 
at the cell interface for obtaining the space derivatives at the interface.
Any reconstruction procedure such as the ENO reconstruction or a fixed-stencil reconstruction
may be used for obtaining these spatial derivatives. 
The `fixed-stencil' corresponds to higher-order polynomial reconstruction using 
constant stencil of FV cells, as against dynamic stencil selection in the ENO reconstruction.
For a $1$D linear system, $k^{th}$ DRP is given by, 

\begin{equation}\label{DRPk}
	\partial_t(\partial_x^{k}{\bf U}) + {\hat A} \partial_x(\partial_x^{k}{\bf U}) = {\bf 0}
\end{equation}

where, 

\begin{equation*}
	\partial_x^{k}{\bf U}(x,0) = \left\{ \begin{array}{lr}
			\partial_x^{k}{\bf U_L}(0) &\hspace{10mm} \text{if} \hspace{5mm} x < 0,  \\
			\partial_x^{k}{\bf U_R}(0) &\hspace{10mm} \text{if} \hspace{5mm} x > 0.  
		\end{array}
		\right.
\end{equation*}

For linear systems, the coefficient matrix ${\hat A}$ is constant. The direction of 
characteristic waves at the 
interface is found out only once during computation of the Godunov state. 
The same information can be used when solving for higher-order derivatives.
Further details regarding the ADER procedure and $2$D implementation
can be found in Ref.\cite{Schwartzkopff2002}.
Once the Taylor series approximation to the state variable vector at the face is evaluated, 
the numerical flux can be computed as 
${\mathcal F} = {\hat A} {\bf\tilde U}$, where  ${\hat A}$ is 
the constant coefficient matrix and ${\bf \tilde U}$ is the Taylor approximation to the state variable vector 
at the cell interface obtained by solving equation\eqref{UR}.
This ADER flux ${\mathcal F}$ is in turn used in the evolution equation\eqref{eq:ode} for 
updating the solution. The update formula using Euler's time-step reads,
\begin{equation}\label{eq:ader}
	{\bf U}_i^{(n+1)}= {\bf U}_i^{(n)} + {\Delta t} \mathcal{R}_{i,p}^{(n)}
\end{equation}
where, ${\Delta t}$ is the time-step governed by the CFL condition.
The numerical scheme given by equation\eqref{eq:ader} is higher-order accurate both in space and 
time as a result of spatial-temporal coupling in the Lax-Wendroff procedure. 
The ADER scheme coupled with the ML method (ML-ADER) can be summarized as follows
(note: since computations in $i^{th}$ FV cell are considered, subscript $i$ is dropped everywhere),
\begin{itemize}
	\item Define number of levels in each cycle ($m$) and total number of cycles ($N$)
	\item At the highest level ($p=m$), Compute $\mathcal{R}_p^{(0)}$ and 
		$\mathcal{R}_{p-1}^{(0)}$ using ADER fluxes and set $\tau_m^{(0)} = 0$ in all FVs.
	\item Compute $\tau_{m-1}^{(1)} = [\mathcal{R}_{p=m}^{(0)} - \mathcal{R}_{p-1}^{(0)} ]$ 
	\item Advance to time-level $1$ by updating the state variable by numerically integrating
		$$\frac{dU}{dt} = \mathcal{R}_{p=m}^{(0)} + \tau_{m}^{(0)}$$. 
	\item At time-level $1$, $p=m-1$. Compute $\mathcal{R}_{p=m-1}^{(1)}$ and $\mathcal{R}_{p-1}^{(1)}$
	\item Compute $\tau_{m-2}^{(2)} = [\mathcal{R}_{p=m-1}^{(1)} - \mathcal{R}_{p-1}^{(1)} ] + \tau_{m-1}^{(1)}$ 
	\item Advance to time-level $2$ by updating state variable by solving $$\frac{dU}{dt} = \mathcal{R}_{p}^{(1)} + \tau_{m-1}^{(1)}$$ 
	\item Continue recursively till the lowest order 
	\item At the lowest order, $p=1$,
		\begin{itemize}
			\item Continue adding $\tau_{0}^{(m)}$ at the lowest level for required number of frozen time-steps ($\nu$). Note that for a sawtooth cycle, $\nu = 1$.
			\item Return to the highest-order accurate operator
		\end{itemize}
	\item This finishes $1$ ML cycle. Continue the process for $N$ cycles.
\end{itemize}
Fig.\ref{mlbetter} shows the ML method in a saw-tooth form, in which a single time step is 
executed at all levels of approximation. 
Transition from the lowest order ($1$) to the highest order ($m$)
does not require the forcing function to be added.

\subsection{Scalar Law}
In this section, we derive conditions under which the modified residual used at the lower level 
approximation can be expected to be higher-order accurate.
Consider the linear advection equation on a $1$D domain, 
\begin{equation}\label{adv}
u_t + au_x = 0, \hspace{4mm} a>0 
\end{equation}
In order to solve the evolution equation (\ref{eq:ader}), 
numerical fluxes ($f = au$) at each face of all finite volumes are required to be computed.
For the $p^{th}$ order ADER method, the inter-cell flux is given as, $$f_{i+\frac{1}{2},r}^{(n)} = a \tilde u(x_{i+\frac{1}{2}},t_n)$$
Where, $\tilde u(x_{i+\frac{1}{2}},t_n)$ is the Taylor series approximation to the state variable $u(x,t)$ at face $i+\frac{1}{2}$
at time level $n$.
In case of the upwind flux (considering $a>0$) $\tilde u(x_{i+\frac{1}{2}},t_n)$ is written as,
\begin{equation}
	\tilde u(x_{i+\frac{1}{2}},t_n) = \sum_{k=0}^{p-1} \frac{(-1)^k a^k (\Delta t)^k}{(k+1)!} \frac{d^k}{dx^k} q^{(n)}_{-}(x_{i+\frac{1}{2}})
\end{equation}
where, $q^{(n)}_{-}(x_{i+\frac{1}{2}})$ is the {$(p-1)$}-order accurate reconstruction polynomial 
interpolating value of the state variable from left side of the face at time level $n$. 
The left interpolated value of the variable is used as a result of 
upwinding.
Thus a generic second order flux at time level $n$ using ADER method can be written as, 
\begin{equation}
	f_{i+\frac{1}{2},2}^{(n)} = a \sum_{k=0}^{1} \frac{(-1)^k a^k (\Delta t)^k}{(k+1)!} \frac{d^k}{dx^k} q_{-}^{(n)}(x_{i+\frac{1}{2}})
\end{equation}
For a second order scheme, the reconstruction polynomial using cell averages using cells $i$ and $i+1$
takes the general form,
$$ q^{(n)}(x) = \alpha(x)u_i^{(n)} + \beta(x)u_{i+1}^{(n)} , \hspace{4mm} \alpha + \beta = 1$$
and the first derivative,
$$ \frac{d}{d x}q^{(n)}(x) = \alpha'(x)u_i^{(n)} + \beta'(x)u_{i+1}^{(n)} $$
Relationship in $\alpha$ and $\beta$ directly yields,
$\alpha' = -\beta'$.
Thus the second order upwind flux at time level $n$ is given as,
\begin{equation}\label{o2}
	f_{i+\frac{1}{2},2}^{(n)} = a[\alpha u_i^{(n)} + \beta u_{i+1}^{(n)} - \frac{a\Delta t}{2} (\alpha'u_i^{(n)} + \beta'u_{i+1}^{(n)})] 
\end{equation}
While the first order upwind flux reads as,
\begin{equation}\label{o1}
	f_{i+\frac{1}{2},1}^{(n)} = a u_i^{(n)}
\end{equation}
From equation(\ref{tau}), the relative truncation error based on inter-cell fluxes is defined as,
\begin{equation*}\label{}
	\tau_{i+\frac{1}{2}}^{(n+1)} = f_{{i+\frac{1}{2}},2}^{(n)} - f_{{i+\frac{1}{2}},1}^{(n)}
\end{equation*}
i.e.
\begin{equation}\label{}
	\tau_{i+\frac{1}{2}}^{(n+1)} = a[\alpha u_i^{(n)} + \beta u_{i+1}^{(n)} - \frac{a\Delta t}{2} (\alpha'u_i^{(n)} + \beta'u_{i+1}^{(n)}) - u_i^{(n)}] 
\end{equation}
This relative truncation error and the `cell-wise' relative truncation error are related as,
$\tau_{i}^{(n)} = \tau_{i+\frac{1}{2}}^{(n)} + \tau_{i-\frac{1}{2}}^{(n)}  $

Since we focus on face $i+\frac{1}{2}$, subscript $i+\frac{1}{2}$ is dropped everywhere.
Using, the relationship $\alpha + \beta = 1$, the relative truncation error can then be written as, 
\begin{equation}\label{tauF}
	\tau^{(n+1)} = a[\beta (u_{i+1}^{(n)}-u_i^{(n)}) - \frac{a\Delta t}{2} (\alpha'u_i^{(n)} + \beta'u_{i+1}^{(n)})] 
\end{equation}
For a two level method, the modified flux at time level $n+1$ is obtained as,
\begin{equation*}\label{}
	{f}_{mod}^{(n+1)} = f_{1}^{(n+1)} + \tau^{(n+1)}
\end{equation*}
i.e. 
\begin{equation}\label{}
	{f}_{mod}^{(n+1)} =a u_i^{(n+1)} + a[\beta (u_{i+1}^{(n)}-u_i^{(n)}) - \frac{a\Delta t}{2} (\alpha'u_i^{(n)} + \beta'u_{i+1}^{(n)})] 
\end{equation}
This modified flux will be higher-order accurate if,
\begin{equation*}\label{}
	{f}_{mod}^{(n+1)} = f_{2}^{(n+1)} 
\end{equation*}
i.e. if
\begin{equation*}\label{}
	a u_i^{(n+1)} + a[\beta (u_{i+1}^{(n)}-u_i^{(n)}) - \frac{a\Delta t}{2} (\alpha'u_i^{(n)} + \beta'u_{i+1}^{(n)})] = a[\alpha u_i^{(n+1)} + \beta u_{i+1}^{(n+1)} - \frac{a\Delta t}{2} (\alpha'u_i^{(n+1)} + \beta'u_{i+1}^{(n+1)})] 
\end{equation*}

In order to find the condition under which the above equation holds true, 
we write all terms using exact solution of the linear advection equation (\ref{adv}). 
To do so, we reconstruct a piecewise polynomial function $\hat u(x,t)$ defined over $x$
constructed using cell averages $u_i$. Thus $$\hat u(x_i,t_n) = u_i^{(n)}$$.
Then the discrete equation can be written in terms of the polynomial approximation function as follows,
\begin{multline*}
	\hat u(x,t+\Delta t) + \beta [\hat u(x+\Delta x,t)-\hat u(x,t)] - \frac{a\Delta t}{2} (\alpha'\hat u(x,t) + \beta'\hat u(x+\Delta x,t)) =\\ \alpha \hat u(x,t+\Delta t) + \beta \hat u(x+\Delta x,t+\Delta t) - \frac{a\Delta t}{2} (\alpha'\hat u(x,t+\Delta t) + \beta'\hat u(x+\Delta x,t+\Delta t)) 
\end{multline*}
Considering advection along characteristics $x/t=a$, i.e. $\Delta x = a\Delta t$, we can write all terms uniformly at time $t$
\begin{multline*}
	\hat u(x-\Delta x,t) + \beta [\hat u(x+\Delta x,t)-\hat u(x,t)] - \frac{a\Delta t}{2} (\alpha'\hat u(x,t) + \beta'\hat u(x+\Delta x,t)) =\\ \alpha \hat u(x-\Delta x,t) + \beta \hat u(x,t) - \frac{a\Delta t}{2} (\alpha'\hat u(x-\Delta x,t) + \beta'\hat u(x,t)) 
\end{multline*}
Substituting $\beta = 1-\alpha$ and also knowing that
$\alpha' = -\beta'$,  the equation reads as,
\begin{equation*}
	\beta \left(\hat u(x-\Delta x,t) + \hat u(x+\Delta x,t) - 2 \hat u(x,t)\right) = \frac{a \Delta t}{2} \beta' \left(\hat u(x-\Delta x,t) + \hat u(x+\Delta x,t) - 2 \hat u(x,t)\right)
\end{equation*}
For a non-trivial solution $\beta \ne 0$ and hence,
\begin{equation}\label{condition}
	\hat u(x,t) = \frac{\hat u(x+\Delta x,t) + \hat u(x-\Delta x,t)}{2}
\end{equation}
This, in the limit $\Delta x \to 0$ translates to condition of differentiability \cite{Chatterjee2015}.
So for sufficiently smooth data, the modified flux is second order accurate in space. 
The ML method thus retains higher-order spatial accuracy by addition of the truncation error as 
a forcing function if the solution is locally linear. 
It can be shown that for even higher-order accuracy, sufficiently `smooth' data would similarly
formally require existence of appropriate higher-order derivatives.
In this analysis, time-accurate advection is considered.
Effect of time-step is explored in subsequent sections.

\subsection{System of linear hyperbolic PDEs}

Linear hyperbolic PDEs are used for modelling many physical systems, e.g. electromagnetic 
waves and aeroacoustic waves.
A generic $1$D system of equation is given by $$U_t + \hat A U_x = 0$$
where $U$ is the conserved variable vector $U=[u_1,u_2,...,u_N]^T$ and 
$\hat A$ is the coefficient matrix which depends on the system parameters. 
This system is strictly hyperbolic if $\hat A$ has $N$ distinct eigenvalues ($\lambda_q, q\in\{1,...,N\}$)
and corresponding $N$ independent right eigenvectors. In such a case, the system can be decoupled into
$N$ linear advection equations as $$(w_q)_t + \lambda_q (w_q)_x = 0, \hspace{5mm} q=1,2,...,N$$
such that, $$U = \sum_{q=1}^N r_q w_q $$ where, $r_q$ is the $q^{th}$ right-eigenvector of $\hat A$. 
The analysis for scalar law is valid for each one of these decoupled equations. 
Hence the system of hyperbolic PDEs can be solved in eigenvector space using a lower-order
accurate operator along with a correction term to get a higher-order accurate solution. 
If the initial data is sufficiently smooth, then the method will yield highest-order
accurate solution to the linear hyperbolic PDEs at lower cost.

\subsection{Multidimensional Systems}
Consider the two-dimensional system of linear hyperbolic equations given by,
$$ \frac{\partial {\bf U}}{\partial t} + \frac{\partial {\bf f}}{\partial x} + \frac{\partial {\bf g}}{\partial y} = {\bf 0}$$
The $p^{th}$ order finite volume discretization of the above system can be written as \cite{Chatterjee2015}
$$ \frac{d {\bf U}_i}{d t} = R_{i,p}^{(n)} = \frac{-1}{\Omega_i} \sum_{j=1}^{k} {[(\mathcal F.\hat n {\mathcal S})_j}]_i $$
where, ${\mathcal F} $ is the flux vector with ${\bf f}$ and ${\bf g}$ components in $x$ and $y$
directions respectively. The system is integrated on a arbitrary finite volume cell with volume $\Omega_i$.
$\mathcal S$ is the length of a face of the FV cell and $\hat n$ is the unit vector pointing 
in the outward direction normal to the face.
In the ML method, $p$-coarsening is performed at each consecutive level and the 
local truncation error is added as the forcing function to restore the highest order of accuracy.
Thus, we solve $ \frac{d {\bf U}_i}{d t} =  R_{i,mod}^{(n)} =  R_{i,p}^{(n)} + \tau_{i,p}^{(n)}$.
The ML method can be similarly extended for $3$D simulations \cite{Chatterjee2015}.

\section{Analysis of ML Method}
\subsection{Spatial Accuracy}

For testing the numerical accuracy of the scheme, 
the scalar linear advection equation $$u_t + au_x = 0, \hspace{4mm} x\in[-1,1], \hspace{4mm} a=1$$ is solved 
with periodic boundaries using Single-Level (SL) ADER and the ML-ADER methods. 
A sine distribution $u(x,0) = sin(\pi x)$ is used as the initial condition.
In both the cases, the simulation is run for time $t=1$ units. 
Fixed-stencil reconstruction is used for both ADER and ML-ADER flux computation. 
In both ADER and ML-ADER simulations, time-step is kept sufficiently small so that 
effect of time-stepping is negligible and spatial errors dominate the overall error.
$L_1$-errors, $L_\infty$-errors and corresponding $L_1$-orders, $L_\infty$-orders for 
the traditional ADER method are presented here in table \ref{ADER1}.
In table \ref{ADERML}, $L_1$ and $L_\infty$ errors and corresponding orders for ML-ADER method are shown.
It can be seen that the numerical accuracy of ML-ADER method is comparable to that of traditional ADER scheme.


\begin{table}[h]
\begin{center}
\begin{tabular}{ | l | l | l | c | l | c  | }
	
	\hline 
	\multicolumn{6}{|c|}{\bf Grid convergence for ADER} \\
	\hline 
	\multicolumn{1}{|c|}{Order} & \multicolumn{1}{|c|}{Cells}& \multicolumn{1}{|c|}{$L_1$ error} & \multicolumn{1}{|c|}{$L_1$ order}& \multicolumn{1}{|c|}{$L_\infty$ error} & \multicolumn{1}{|c|}{$L_\infty$ order} \\
	\hline                           
	2 & 25 &	2.10E-02&	-    &3.30E-02 &	-       \\
	  & 50 &	5.26E-03&	2.00 &8.26E-03 &	2.00    \\
	  & 100&	1.32E-03&	2.00 &2.07E-03 &	2.00    \\
	  & 200&	3.30E-04&	1.99 &5.19E-04 &	1.99    \\
	\hline 
	3 & 25 &	2.62E-03&	-    &4.12E-03 &	-       \\
	  & 50 &	3.30E-04&	2.99 &5.18E-04 &	2.99    \\
	  & 100&	4.13E-05&	3.00 &6.49E-05 &	3.00    \\
	  & 200&	5.16E-06&	3.00 &8.11E-06 &	3.00    \\
	\hline 
	4 & 25 &	3.95E-04&	-    &6.21E-04 &	-       \\
	  & 50 &	2.49E-05&	3.99 &3.90E-05 &	3.99    \\
	  & 100&	1.56E-06&	4.00 &2.44E-06 &	4.00    \\
	  & 200&	9.73E-08&	4.00 &1.53E-07 &	4.00    \\
	\hline

\end{tabular}
	\caption{\label{ADER1} Grid convergence study for the ADER scheme with fixed-stencil reconstruction}
\end{center}
\end{table}

\begin{table}[h]
\begin{center}
\begin{tabular}{ | l | l | l | c | l | c  | }
	
	\hline 
	\multicolumn{6}{|c|}{\bf Grid convergence for ML-ADER} \\
	\hline 
	\multicolumn{1}{|c|}{Order} & \multicolumn{1}{|c|}{Cells}& \multicolumn{1}{|c|}{$L_1$ error} & \multicolumn{1}{|c|}{$L_1$ order}& \multicolumn{1}{|c|}{$L_\infty$ error} & \multicolumn{1}{|c|}{$L_\infty$ order} \\
	\hline                           
	2 & 25 &	2.10E-02&	-    &	3.30E-02   &	-       \\
	  & 50 &	5.27E-03&	2.00 &	8.27E-03   &	2.00    \\
	  & 100&	1.32E-03&	2.00 &	2.07E-03   &	2.00    \\
	  & 200&	3.31E-04&	1.99 &	5.20E-04   &	1.99    \\
	\hline 
	3 & 25 &	2.62E-03&	-    &	4.11E-03   &	-       \\
	  & 50 &	3.29E-04&	2.99 &	5.17E-04   &	2.99    \\
	  & 100&	4.13E-05&	3.00 &	6.48E-05   &	3.00    \\
	  & 200&	5.17E-06&	3.00 &	8.12E-06   &	3.00    \\
	\hline 
	4 & 25 &	3.94E-04&	-    &	6.19E-04   &	-       \\
	  & 50 &	2.45E-05&	4.01 &	3.84E-05   &	4.01    \\
	  & 100&	1.46E-06&	4.07 &	2.29E-06   &	4.07    \\
	  & 200&	7.27E-08&	4.33 &	1.14E-07   &	4.33    \\
	\hline 

\end{tabular}
	\caption{\label{ADERML} Grid convergence study for the ML-ADER scheme with fixed-stencil reconstruction}
\end{center}
\end{table}

\begin{table}[h]
\begin{center}
\begin{tabular}{ | l | l | l | c | l | c  | }
	
	\hline 
	\multicolumn{6}{|c|}{\bf Grid convergence with broadband initial conditions} \\
	\hline 
	\multicolumn{1}{|c|}{Theoretical} & \multicolumn{1}{|c|}{Number of}& \multicolumn{2}{|c|}{SL-ADER}& \multicolumn{2}{|c|}{ML-ADER} \\
	\cline{3-4} \cline{5-6} 
	\multicolumn{1}{|c|}{Order} & \multicolumn{1}{|c|}{Cells}& \multicolumn{1}{|c|}{$L_1$ error} & \multicolumn{1}{|c|}{$L_1$ order}& \multicolumn{1}{|c|}{$L_1$ error} & \multicolumn{1}{|c|}{$L_1$ order} \\
	\hline                           
	2 & 25 &	1.01E-04	&0.00 &	1.01E-04	&0.00   \\
	  & 50 &	3.37E-05	&1.59 &	3.37E-05	&1.59   \\
	  & 100&	8.71E-06	&1.95 &	8.71E-06	&1.95   \\
	  & 200&	2.31E-06	&1.91 &	2.31E-06	&1.91   \\
	\hline 
	3 & 25 &	2.20E-04	&0.00 &	2.17E-05	&0.00   \\
	  & 50 &	3.51E-05	&2.65 &	3.46E-06	&2.65   \\
	  & 100&	4.76E-06	&2.89 &	4.79E-07	&2.85   \\
	  & 200&	6.16E-07	&2.95 &	6.20E-08	&2.95   \\
	\hline 
	4 & 25 &	9.81E-06	&0.00 &	9.81E-06	&0.00   \\
	  & 50 &	1.04E-06	&3.24 &	1.04E-06	&3.24   \\
	  & 100&	7.40E-08	&3.81 &	7.40E-08	&3.81   \\
	  & 200&	4.86E-09	&3.93 &	4.86E-09	&3.93   \\
	\hline 

\end{tabular}
	\caption{\label{gaussianord} Grid convergence study with ENO reconstruction for a broadband initial conditions}
\end{center}
\end{table}

In another experiment, a Gaussian initial profile
$$u(x) = e^{(-\tfrac{x}{0.25})^2}, \hspace{4mm} x \in [-1,1]$$ centered at the origin is taken as the initial condition. 
ENO reconstruction is used for both SL-ADER and ML-ADER methods. 
Table \ref{gaussianord} shows L$1$ errors and corresponding orders for SL-ADER and ML-ADER methods.
The ML-ADER method shows similar rate of convergence as the SL-ADER method. 
This study shows that the ML-ADER method yields spatially higher-order accurate 
solution for monochromatic as well as broadband initial conditions. 

\subsection{Effect of Time-Step}
In the ADER class of schemes, temporal accuracy is coupled with the spatial accuracy.
In such a case, the interaction in temporal and spatial errors may depend on the time step \cite{Zingg2000}. 
In this work, we perform a posteriori error analysis in order to study the effect of time-step
on accuracy of the method.
In this analysis, the linear advection equation $u_t + u_x = 0$ is solved
over $x\in[-1,1]$ with periodic boundary conditions. Both conventional (single level) and ML methods
are used for solving this equation. A Gaussian distribution
$$u(x) = e^{(-\tfrac{x}{0.25})^2}, \hspace{4mm} x \in [-1,1]$$ centered at the origin is taken as the initial condition. 
After time $t=2$ units, the signal re-centers at the origin. 
At this point, the signal representation in the frequency domain is obtained using a 
Discrete Fourier Transform (DFT). 
The Fast Fourier Transform (FFT) method is used for obtaining DFT of the solution.
The idea of using DFT for examining dispersion relation of the modified equation resulting from a 
particular numerical scheme is similarly used in the 
Approximate-Dispersion-Relation (ADR) method \cite{Pirozzoli2006}.
If the function $f(x):{\mathbb R} \rightarrow {\mathbb R} $ is sampled at $N$ distinct points \{$x_0,...,x_N$\},
then the DFT of the signal can be written as
$$ \hat f(k) = \frac{1}{N} \sum_{n=0}^{N-1} f(x_n)e^{- i k n}$$ where $k\in {\mathbb{R}}$ is 
the wavenumber of a Fourier mode.
After performing the FFT operation, the Power-Spectral-Density (PSD) curve is obtained.
The PSD is a plot of the energy of a Fourier mode $| \hat f(k) |^2 $ with respect to the
wavenumber $k = \frac{2\pi}{\lambda}$, where $\lambda$ is the corresponding wavelength.

We use the PSD plot for understanding the behavior of errors due to a particular numerical 
approximation. 
The value of the zeroth Fourier mode indicates average value of the signal in the physical domain.
This stems from the conservative Finite-Volume (FV) framework. 
Larger contribution of the lower wavenumbers indicate smooth flow-features in the physical domain.
Therefore, compression of the PSD curve can indicate addition of 
diffusion in the evolution process.
Moreover, spurious higher-wavenumbers in the FFT signal may indicate steep slopes arising due to 
instabilities or nonlinear phenomena. 
Thus, the Fourier representation of the signal  
presents a convenient tool for posteriori error analysis of an upwind numerical scheme.

\begin{figure}[h]
	\begin{minipage}[t]{0.5 \textwidth}
		\begin{center}{\includegraphics[width=0.9\textwidth]{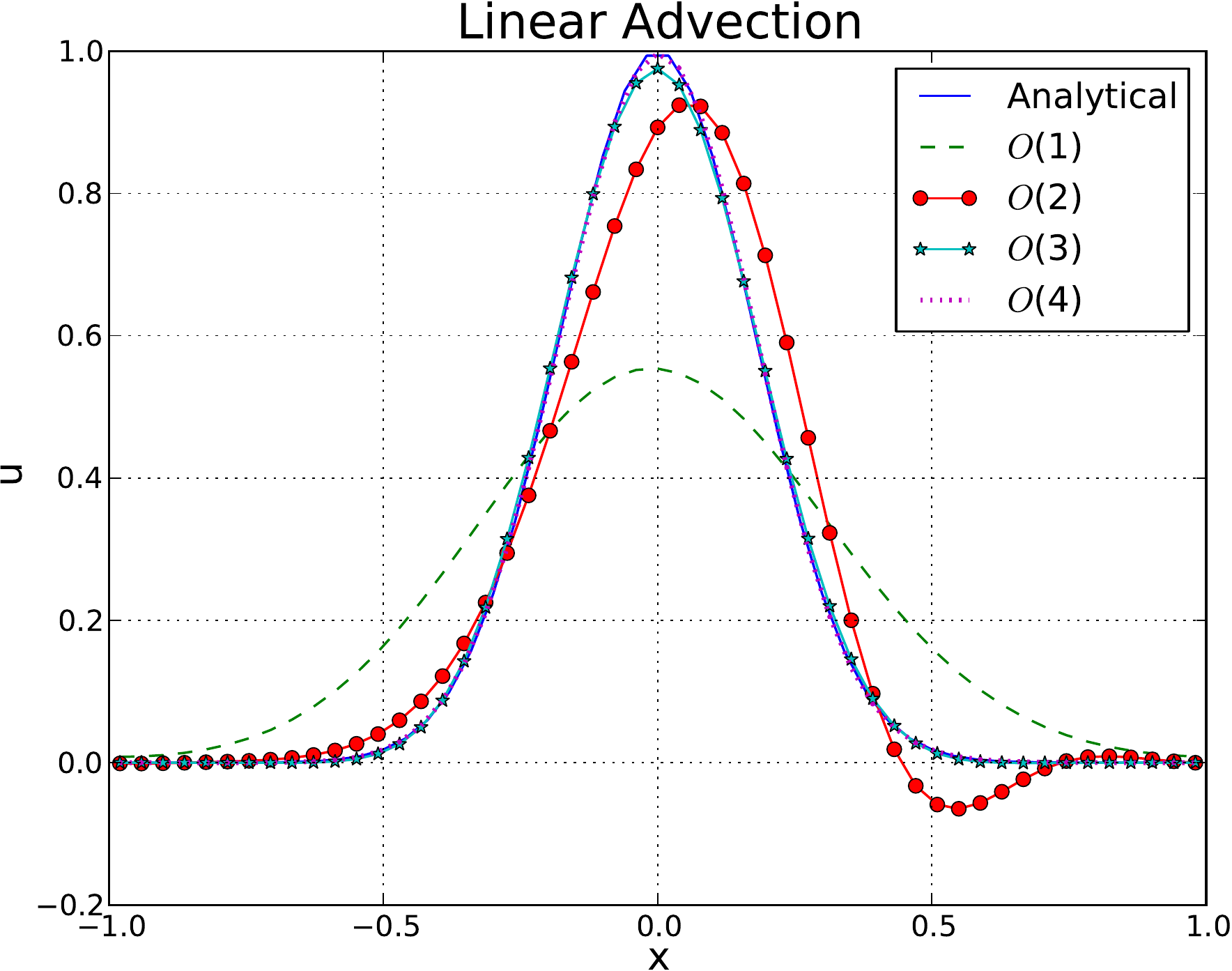}}
		\end{center}
	\end{minipage}%
	\begin{minipage}[t]{0.5 \textwidth}
		\begin{center}{\includegraphics[width=0.9\textwidth]{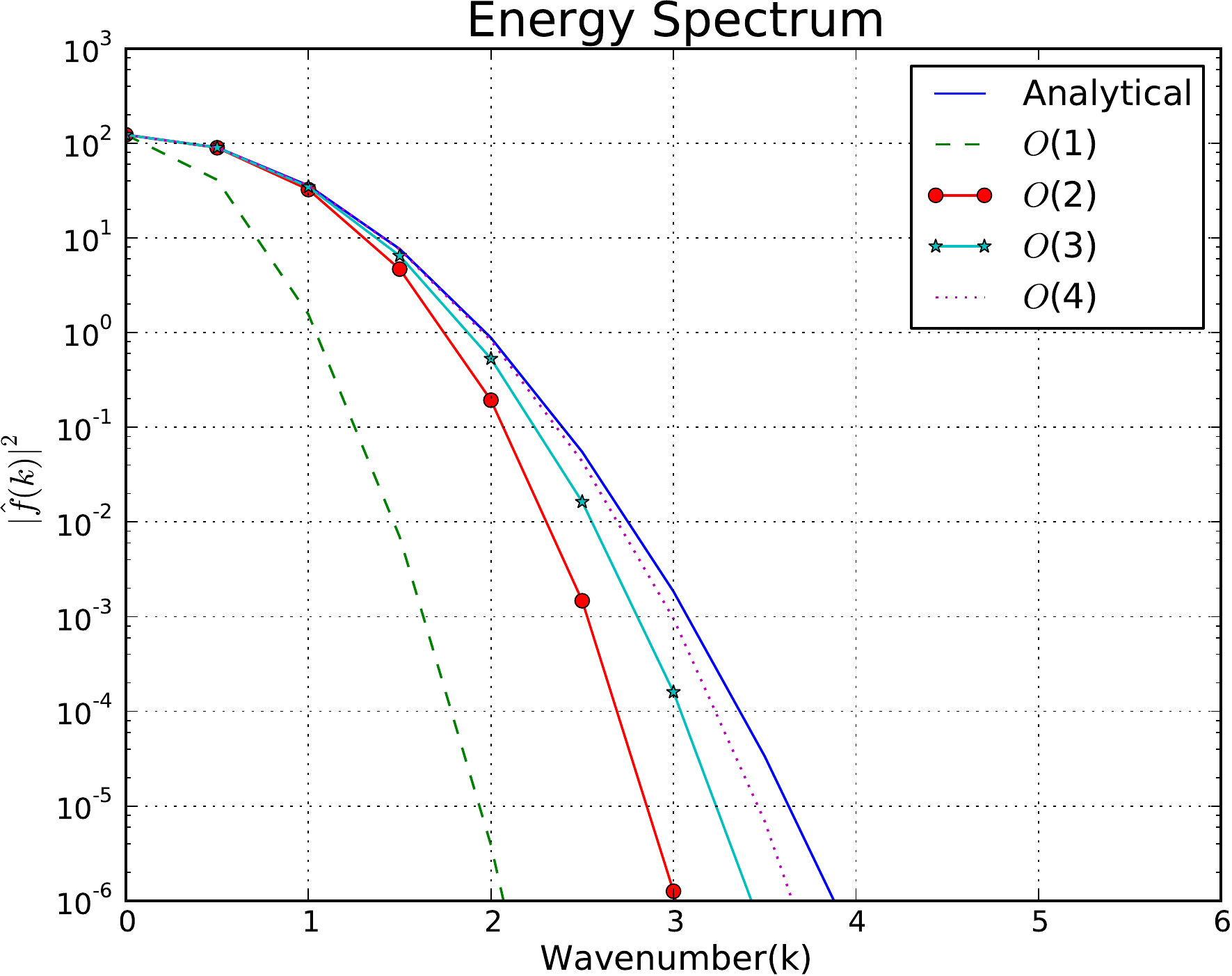}}
		\end{center}
	\end{minipage}%
	\caption{\label{SL}a. Linear wave propagation with fixed-stencil higher-order ADER schemes \hspace{2mm} b. Power spectrum}
\end{figure}

\begin{figure}[h]
	\begin{minipage}[t]{0.47 \textwidth}
		\begin{center}{\includegraphics[width=\textwidth]{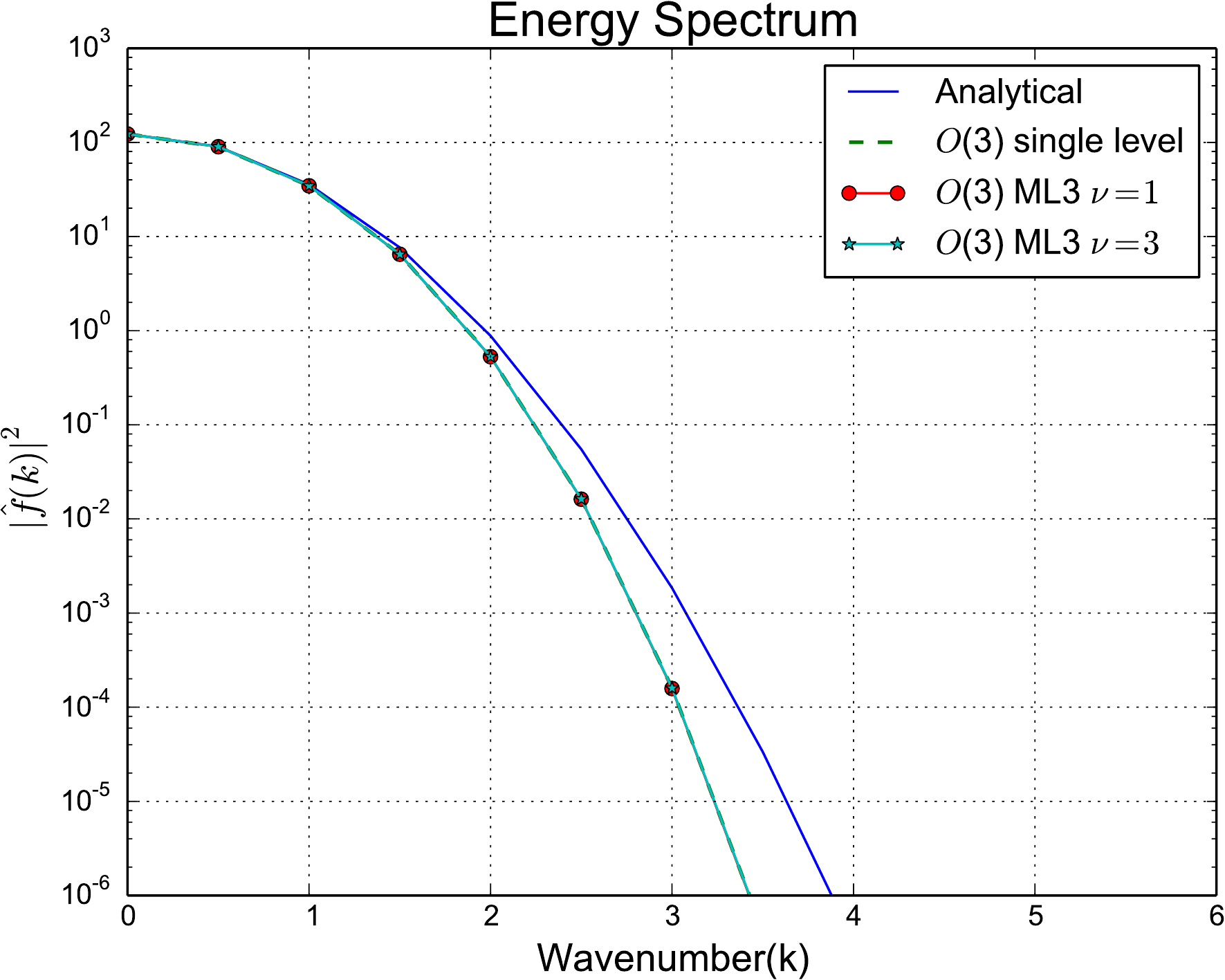}}
		\end{center}
	\end{minipage}%
	\begin{minipage}[t]{0.47 \textwidth}
		\begin{center}{\includegraphics[width=\textwidth]{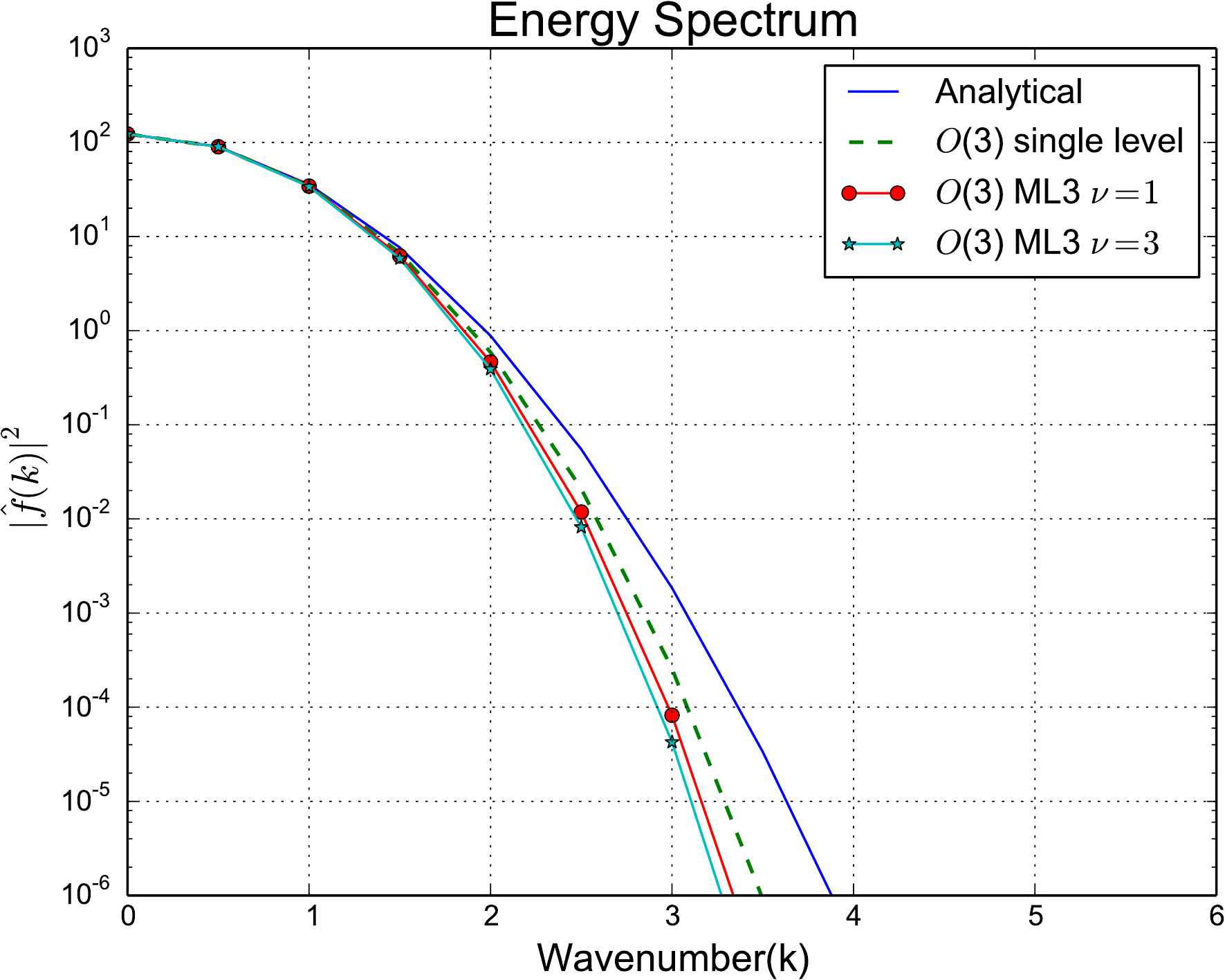}}
		\end{center}
	\end{minipage}%
	\caption{\label{tstep} Power spectrum of results using a. CFL 0.003 (small time step) \hspace{2mm} b. CFL 0.3 (Larger time step) for a third-order accurate scheme }
\end{figure}
\begin{figure}[h]
	\begin{minipage}[t]{0.47 \textwidth}
		\begin{center}{\includegraphics[width=\textwidth]{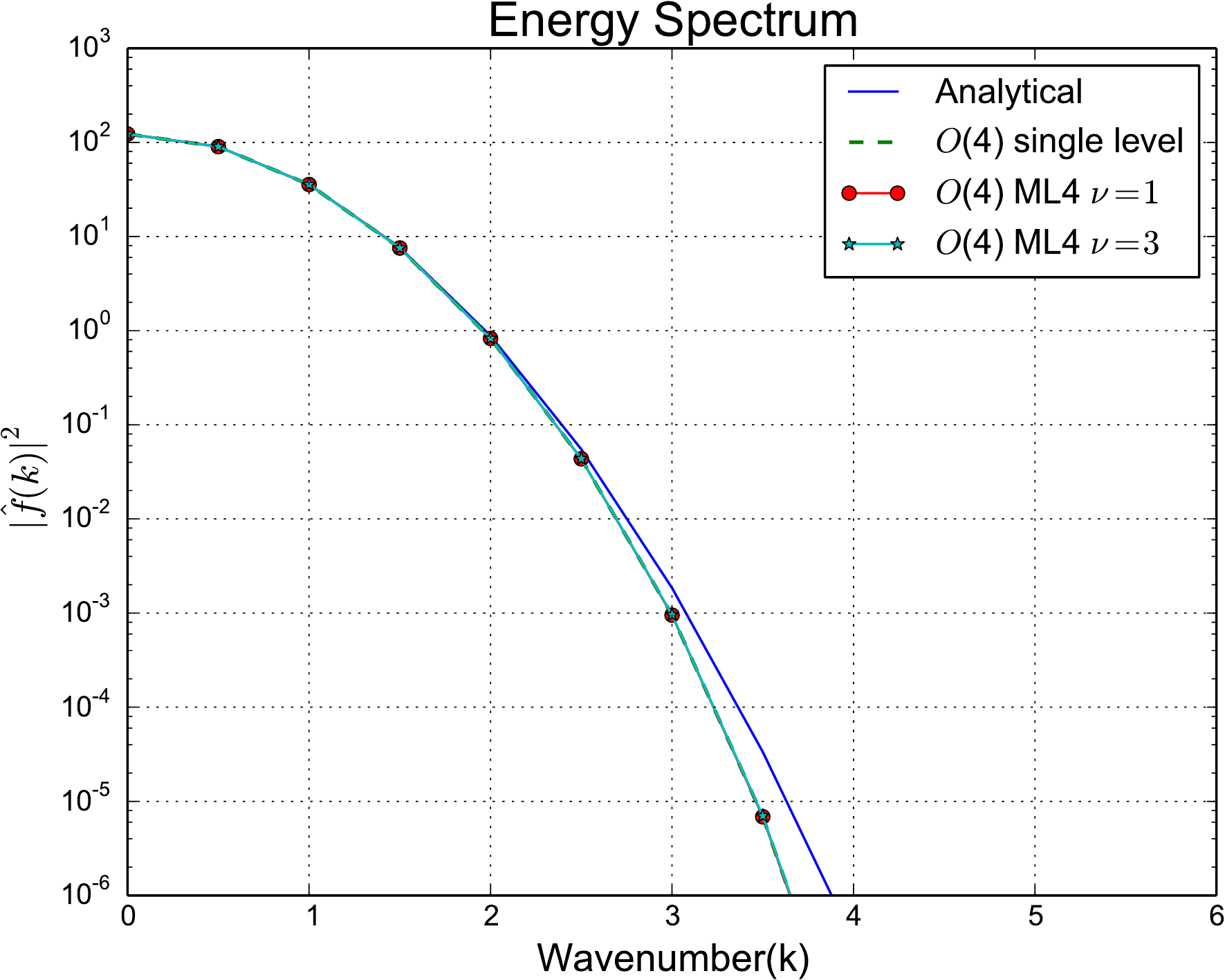}}
		\end{center}
	\end{minipage}%
	\begin{minipage}[t]{0.47 \textwidth}
		\begin{center}{\includegraphics[width=\textwidth]{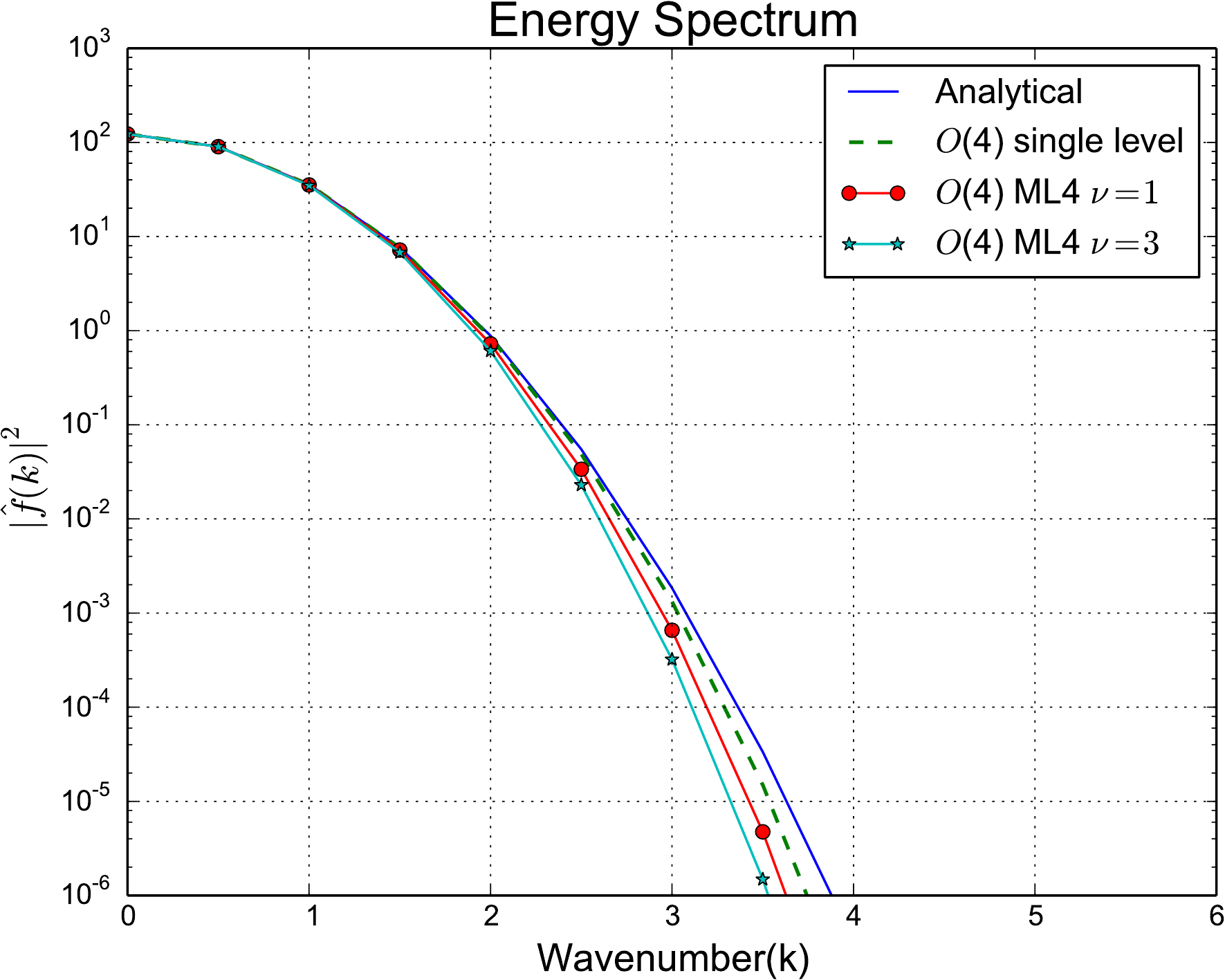}}
		\end{center}
	\end{minipage}%
	\caption{\label{tstep2} Power spectrum of results using a. CFL 0.003 (small time step) \hspace{2mm} b. CFL 0.3 (Larger time step) for a fourth-order accurate scheme }
\end{figure}

We examine effect of the time-step on both saw-tooth and the frozen-$\tau$ cycling of the solution.
In all cases, fixed-stencil
finite volume ADER schemes are used.
An upwind stencil is chosen for the second-order
approximation since the central scheme turns out to be unstable.
Figure \ref{SL}b shows PSD of results computed using conventional 
ADER schemes of varying orders shown in  Fig. \ref{SL}a.
It can be noted that all schemes are conservative. Lower-order schemes show  
mainly lower wavenumbers contributing to the energy spectrum and thus indicating a more diffusive 
solution. 
However, it is to be noted that non dissipative Finite Difference (FD) or FV schemes such as the
central schemes show zero diffusion error but a non-zero dispersion error. 
Analysis presented here
is valid only for upwind numerical schemes.

Fig. \ref{tstep}a and \ref{tstep2}a show the PSD curve corresponding to a very small time-step
for a third-order and a fourth-order schemes respectively. In both cases, the ML method preserves
higher-order spatial accuracy and does not introduce errors
(compression of the PSD curve or presence of spurious higher modes) in the solution.
Also, the ML-ADER method is seen to be conservative in nature.  
When a larger time-step is chosen, the ML-ADER method indicates addition of diffusion in the 
much larger wavenumber components as seen in Fig. \ref{tstep}b and \ref{tstep2}b. 
With multiple frozen steps at the lowest-order approximation (frozen-$\tau$ method), 
more diffusion is progressively added in the larger wavenumber components. 

However, the loss of energy in the larger Fourier modes is much smaller compared to the 
average energy of the zeroth Fourier coefficient $|\hat f(0)|^2$. 
Moreover, as pointed out in Ref.\cite{Zingg2000}, computational cost varies only linearly with the
time-step and thus an economical higher-order spatial approximation is much more beneficial
compared to a large time-step in this context.

\subsection{Effect of Time-Step on Group Velocity}

Group velocity is the velocity with which a wave-envelope travels through dispersive media.
It is understood that the energy in the signal travels at the group velocity \cite{Beta2010,leveque}.
A numerical scheme resulting from finite dimensional approximation of the original PDE 
may have a different group velocity than the theoretical value of group-velocity. 
Ability of numerical schemes in preserving
the shape as well as propagating wave-packets with correct velocity is also
of crucial importance for accurate simulations \cite{Beta2010}.
For evaluating effect on group velocity,
we consider the traveling wave-packet problem presented in Ref.\cite{Beta2010}. 
Although the group-velocity analysis is directly applicable to only non-dissipative numerical 
schemes, for low frequency waves, dispersion dominates diffusion and
thus the analysis proves useful even in the presence of diffusion\cite{Beta2010}. 

\begin{figure}[h]
	\begin{center}
	\includegraphics [scale=0.55]{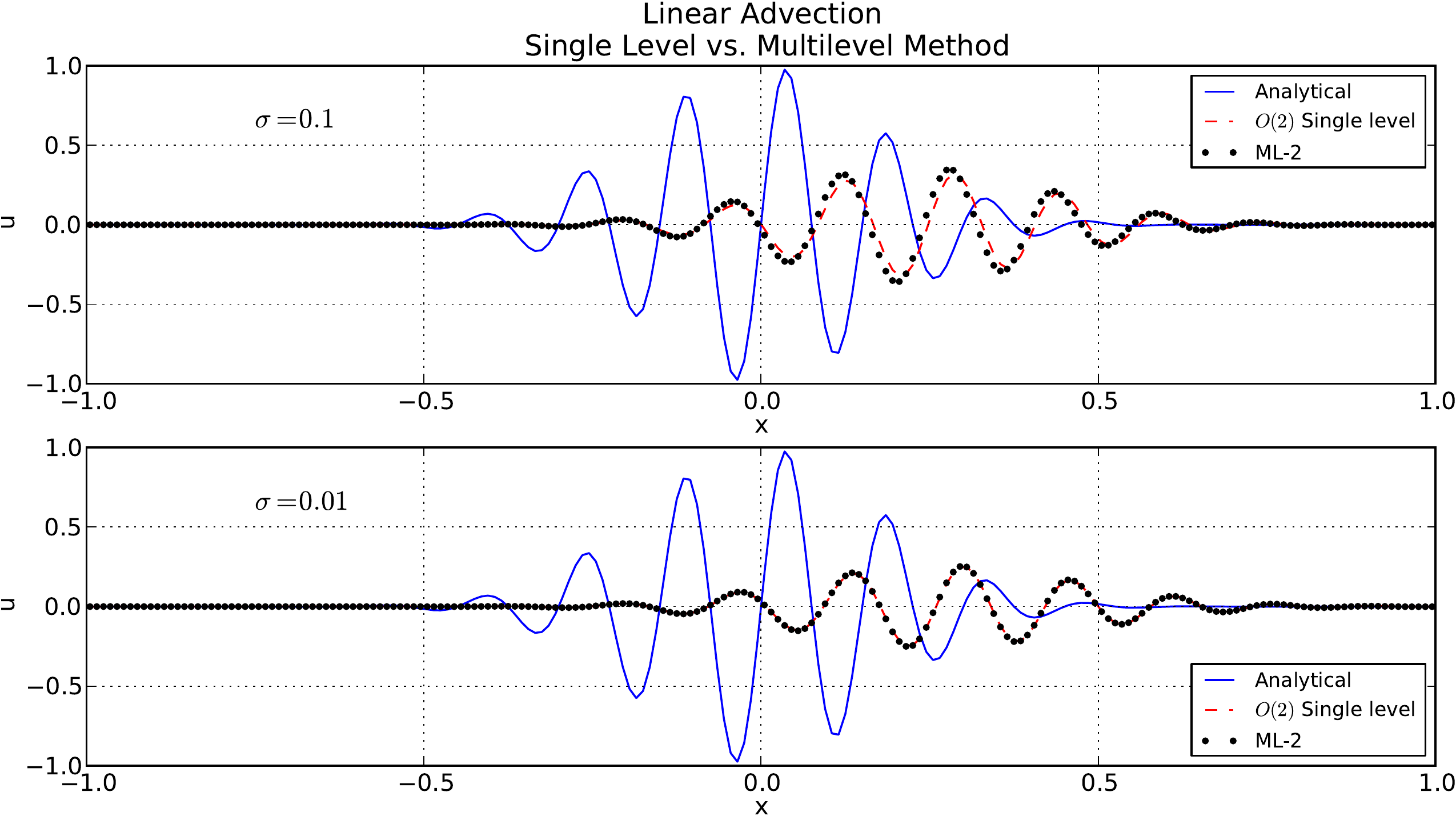}
\end{center}
	\caption{\label{g1} \small{Comparison of second-order ADER scheme with 2-level ADER method a.Large time-step b.Small time-step}}
\end{figure}

%

\begin{figure}[H]
	\begin{center}
	\includegraphics [scale=0.55]{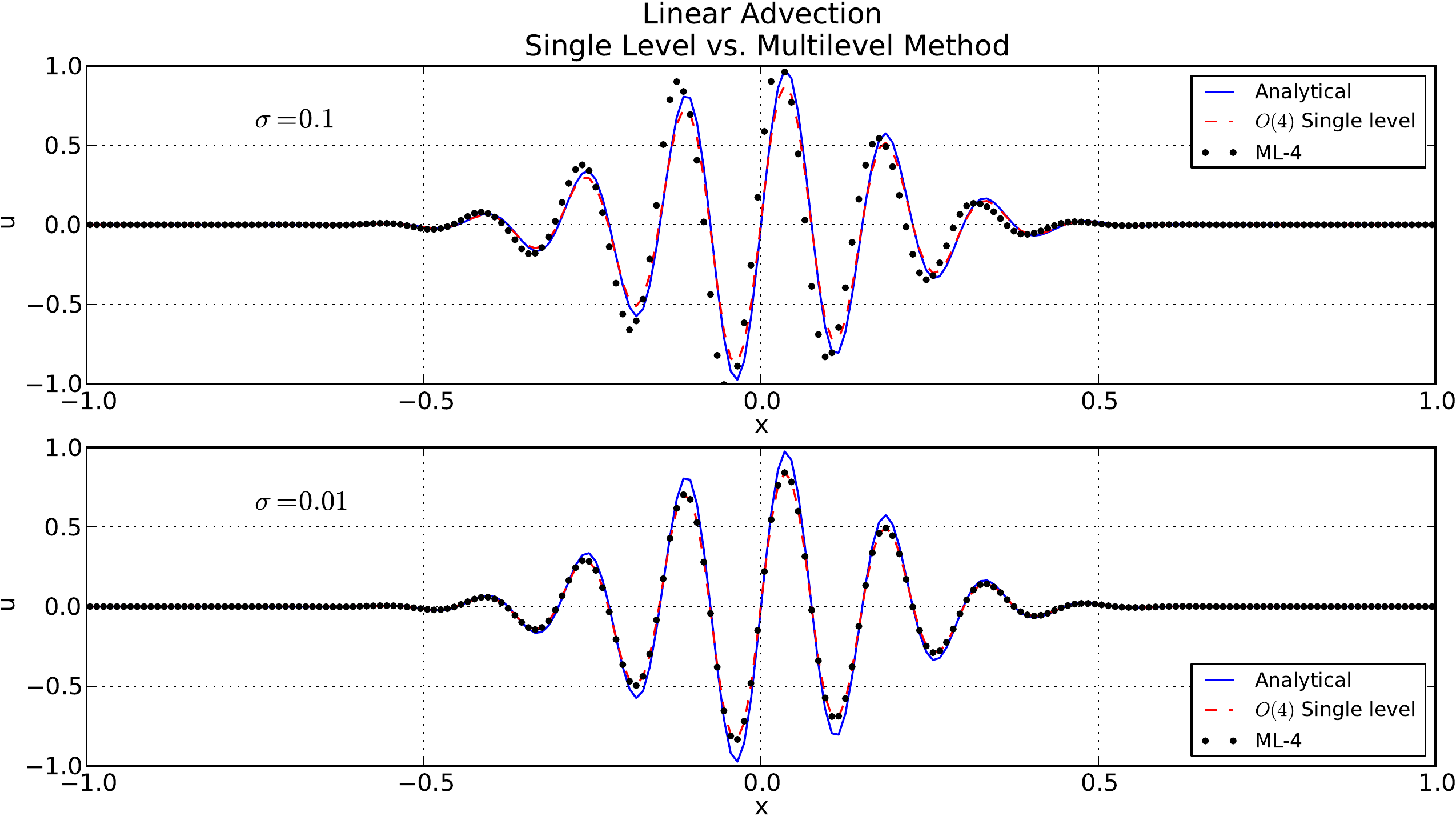}
\end{center}
	\caption{\label{g3} \small{Comparison of fourth-order ADER scheme with 4-level ADER method a.Large time-step b.Small time-step}}
\end{figure}

For this analysis, the linear advection equation given by
equation\eqref{adv} is solved with wave speed $a=1$.
A wave-packet centered at $0$ is given by,
$$u(x,0) = e^{-16(x)^2} sin (\xi x), \hspace{4mm} x\in[-1,1]$$
$\xi$ is the wavenumber corresponding to $15$ PPW so as to concentrate on lower frequency waves. 
Theoretically the wave-packet is expected to re-center at $0$ at time 
$t=2$ units due to periodic boundary conditions. 
However, different numerical schemes have different dispersion characteristics and thus  
transport the wave-packet with different speeds.
We use a second-order, a third-order and a fourth-order ADER schemes as well as the ML variants.
For the second order scheme, we use a fixed upwind stencil for reconstruction, while the
third and the fourth order schemes are constructed with ENO reconstruction.
It may be noted that the ENO or upwind reconstructions used in this study result in
dissipative schemes. 
Dissipative schemes fail to preserve wave amplitude for lower Courant numbers for
a lower PPW.
As the analysis is recommended for low frequency signals in upwind based methods, 
we consider a larger PPW in contrast to $8$ PPW prescribed in Ref.\cite{Beta2010}.
In each case, two different time-steps are used corresponding to the Courant number $\sigma = 0.1$ 
and $\sigma = 0.01$.

Fig.\ref{g1} and \ref{g3} show single-level and multilevel results for 
a second-order and a fourth-order schemes respectively. 
For a small time-step, both ADER and ML-ADER show identical group velocity.
For a larger time-step however, small difference in the group velocity is seen especially at a higher
order.

\section{Application to CAA}\label{caa}

The solution of two problems in computational aeroacoustics is presented using the ML-ADER method.
The first problem \cite{CAA3} deals with propagation of acoustic waves in 
the upstream direction though a near-transonic convergent-divergent nozzle. 
The quasi-$1$D nozzle is described with the following area distribution,
\begin{equation}\label{noz}
	A(x) = \left\{ \begin{array}{lr}
			0.536572 - 0.198086 \ e^{-ln(2)\left(\tfrac{x}{0.6}\right)^2} &\text{,} \hspace{5mm} x > 0,  \\
			1 - 0.661514 \ e^{-ln(2)\left(\tfrac{x}{0.6}\right)^2} &\text{,} \hspace{4mm} x < 0
		\end{array}
		\right.
\end{equation}
, $x \in [-10,10]$.
In order to simulate the acoustic wave propagation, linearized Euler equations in $1$D
\begin{equation}\label{euler1d}
	{\bf U}_t + {\bf A} {\bf U}_x = S
\end{equation}
are solved. ${\bf U}$ is the vector of the acoustic perturbations and ${\bf A}$ is the coefficient
matrix based on the background flow in the nozzle, given as,

\begin{equation}
\mbox{\bf U}\!=\!\left[\begin{array}{c}
\rho\\ u \\ p 
\end{array} \right]\: , \:
\mbox{\bf A}\!=\!\left[\begin{array}{c c c}
		u_0 & \rho_0 & 0 \\ 0 & u_0 & \frac{1}{\rho_0} \\ 0 & \gamma p_0  & u_0 
\end{array} \right]
\label{eq:caa}
\end{equation}
$\rho$, $u$ and $p$ are density, velocity and pressure perturbation-variables respectively.
The steady state values of velocity($u_0$), density($\rho_0$) and pressure($p_0$)
correspond to steady state flow through a 
converging-diverging nozzle. These values can be found out either analytically
or numerically. Analytical expressions for obtaining steady-state values are given in Ref.\cite{CAA3}.
The steady state value of Mach number at exit of the nozzle is $0.4$.
A pulsating acoustic source is situated at the nozzle exit.
The source term is given as, 
\begin{equation}\label{source}
	S = \left[\begin{array}{c} \rho \\ u \\ p \end{array}\right]_{acoustic} = 
	10^{-5} \left[\begin{array}{c} 1 \\ -1 \\ 1 \end{array}\right] 
	cos \left[\omega ( \frac{x}{1-M} + t)\right]
\end{equation}

\begin{figure}[h]
	\begin{center}
	\includegraphics [scale=0.6]{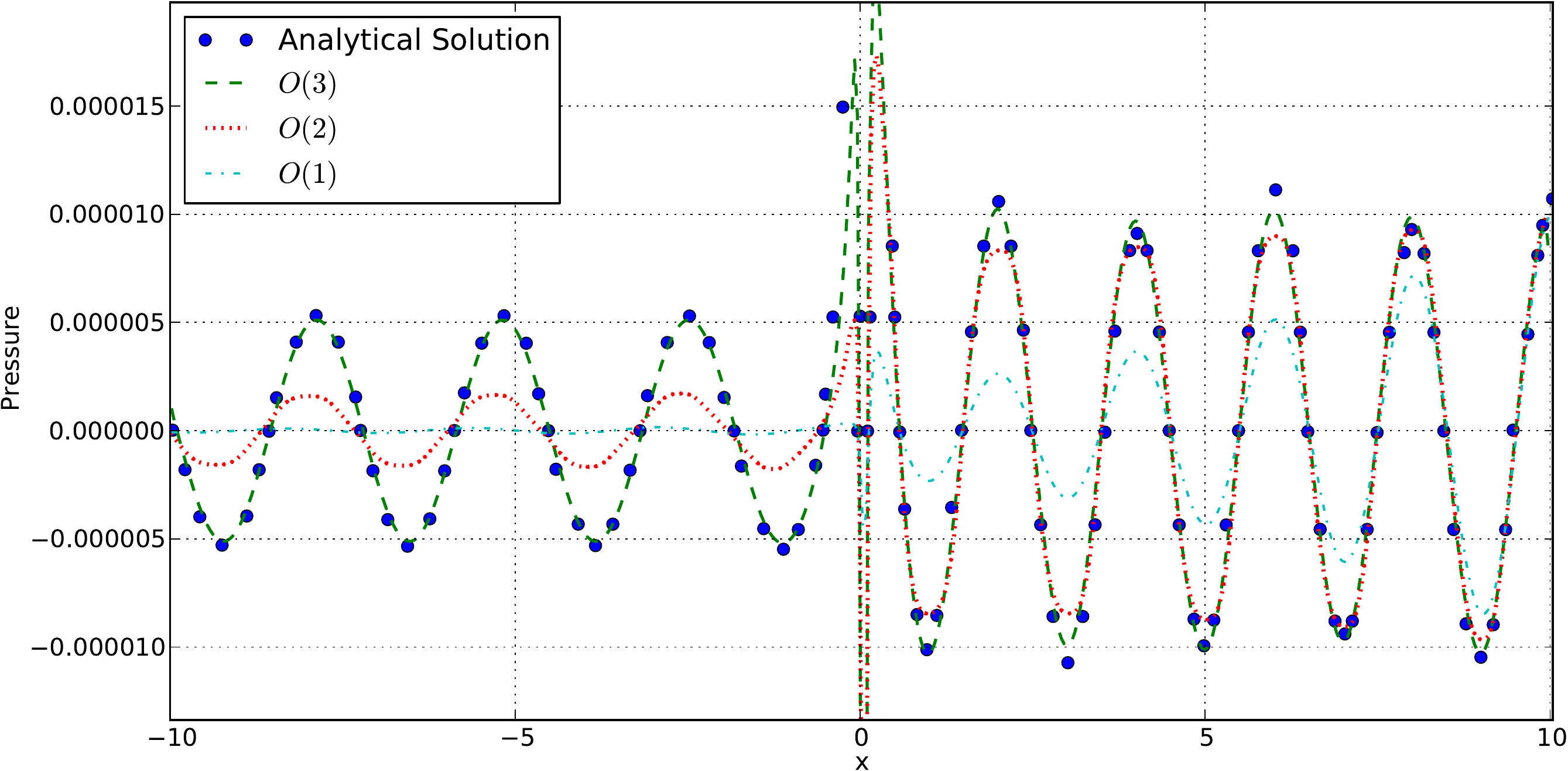}
\end{center}
\caption{\label{Fig1} \small{Acoustic waves travelling in upstream direction in a subsonic nozzle: Comparison of different spatial orders of accuracy.  }}
\end{figure}

\begin{figure}[h]
	\begin{center}
	\includegraphics [scale=0.6]{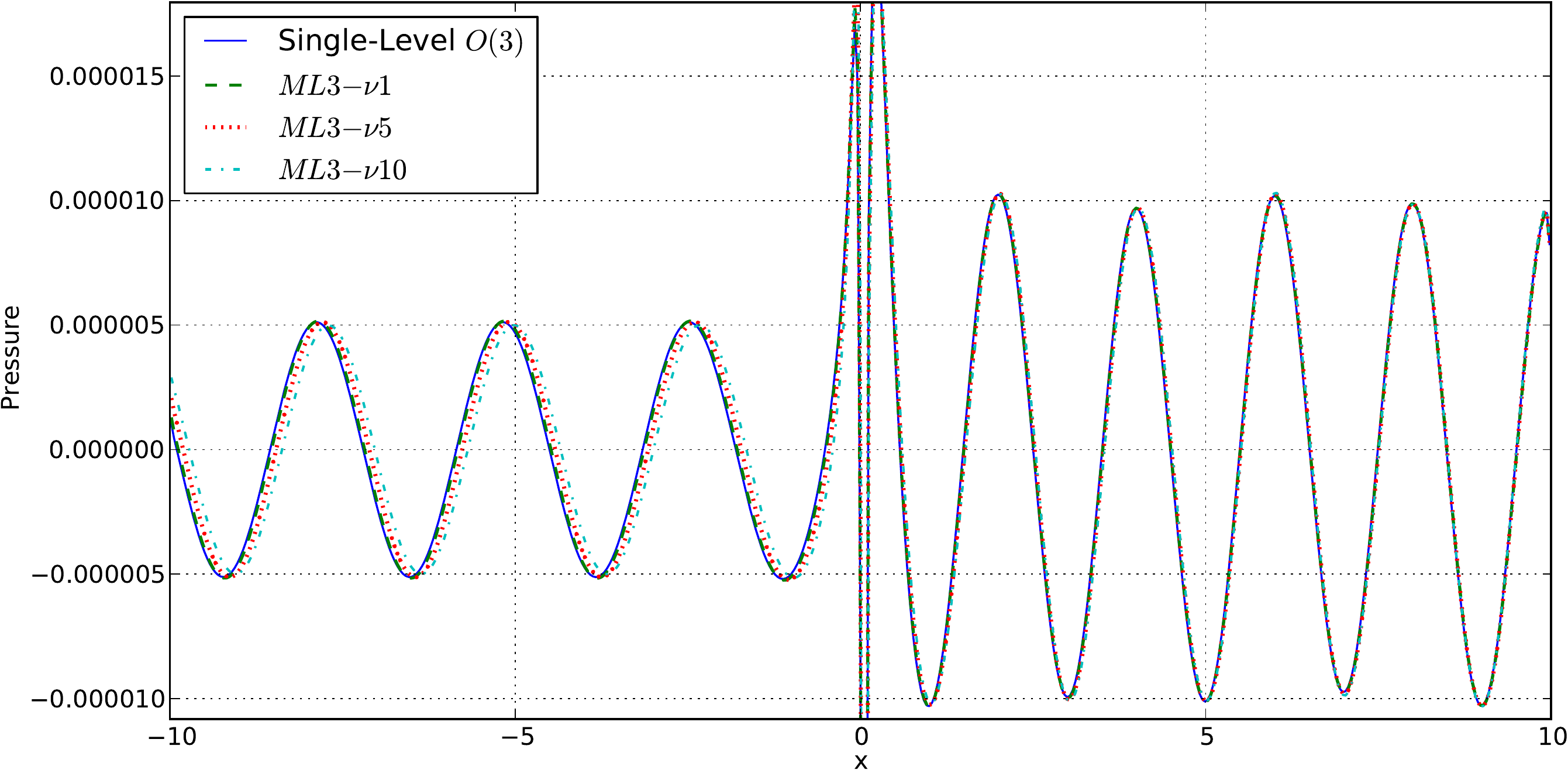}
\end{center}
\caption{\label{Fig2} \small{Acoustic waves travelling in upstream direction in a subsonic nozzle: Comparison of conventional and multilevel $3^{rd}$-order ADER methods }}
\end{figure}

As the flow in the nozzle is subsonic, acoustic waves travel
in the upstream direction through the nozzle. 
For computing the acoustic waves, the system of equations \eqref{euler1d}
is first decoupled into characteristic variables. This is followed
by higher-order reconstruction of state-variables and derivatives which in turn are used for 
ADER flux computations.
A third-order scheme is used for solving this problem, because for the present wave-resolution
a fourth-order scheme yields visually identical results as a third-order scheme. 
Fig.\ref{Fig1} shows acoustic pressure field computed with single-level ADER methods of different
spatial orders of accuracy. For these computations, the $1$D domain is divided into $501$ equi-spaced 
FV cells. It is clear from the figure, that higher order methods result in superior 
resolution of waves.   
Fig.\ref{Fig2} shows comparison of acoustic pressure field computed with 
conventional and multi-level ADER methods. 
Results are obtained with saw-tooth and frozen-$\tau$ variants of ML-ADER method. A
good agreement is seen between single-level and multi-level results. 
It is to be noted that, in this simulation the time step corresponds to $\sigma = 0.25$
and not equal to the small time-step used in the accuracy analysis. 
For multidimensional system of equations, the minor deviation between the SL and the ML
solutions at larger time steps seen for a scalar law becomes insignificant.
The ML method produces comparable results as the conventional method
at reasonable CFL numbers for practical problems.

In the second example \cite{CAA2}, acoustic scattering 
from a $2$D circular cylinder is simulated using conventional and multi-level
ADER methods.
For this simulation, a $2$D system of linearized Euler equations 
\begin{equation}\label{euler2d}
	{\bf U}_t + {\bf A} {\bf U}_x  + {\bf B} {\bf U}_y = 0
\end{equation}is solved.
${\bf U}$ is the vector of the acoustic perturbations and ${\bf A, B}$ are the coefficient
matrices, given as,

\begin{equation}
\mbox{\bf U}\!=\!\left[\begin{array}{c}
p \\ u \\ v 
\end{array} \right]\: , \:
\mbox{\bf A}\!=\!\left[\begin{array}{c c c}
		u_0 & \rho_0 c_0^2 & 0 \\ \frac{1}{\rho_0} & u_0 & 0 \\ 0 & 0  & u_0 
\end{array} \right] , \:
\mbox{\bf B}\!=\!\left[\begin{array}{c c c}
		v_0 & 0 & \rho_0 c_0^2  \\ 0 & v_0 & 0 \\\frac{1}{\rho_0}  & 0  & v_0 
\end{array} \right]
\label{eq:caa2}
\end{equation}
where,
$u,v$ are velocity-perturbation components in $x$ and $y$ direction and 
$p$ is the pressure perturbation.
Initially (at $t=0$)  $u=v=0$ and 
\begin{equation}
	p = exp\left[-ln 2 \left(\frac{(x-4)^2 + y^2}{(0.2)^2}\right)\right]
\end{equation}
with the background velocity, $u_0 = v_0 = 0$.

\begin{figure}[h]
	\begin{center}
		\includegraphics [width=0.9\textwidth]{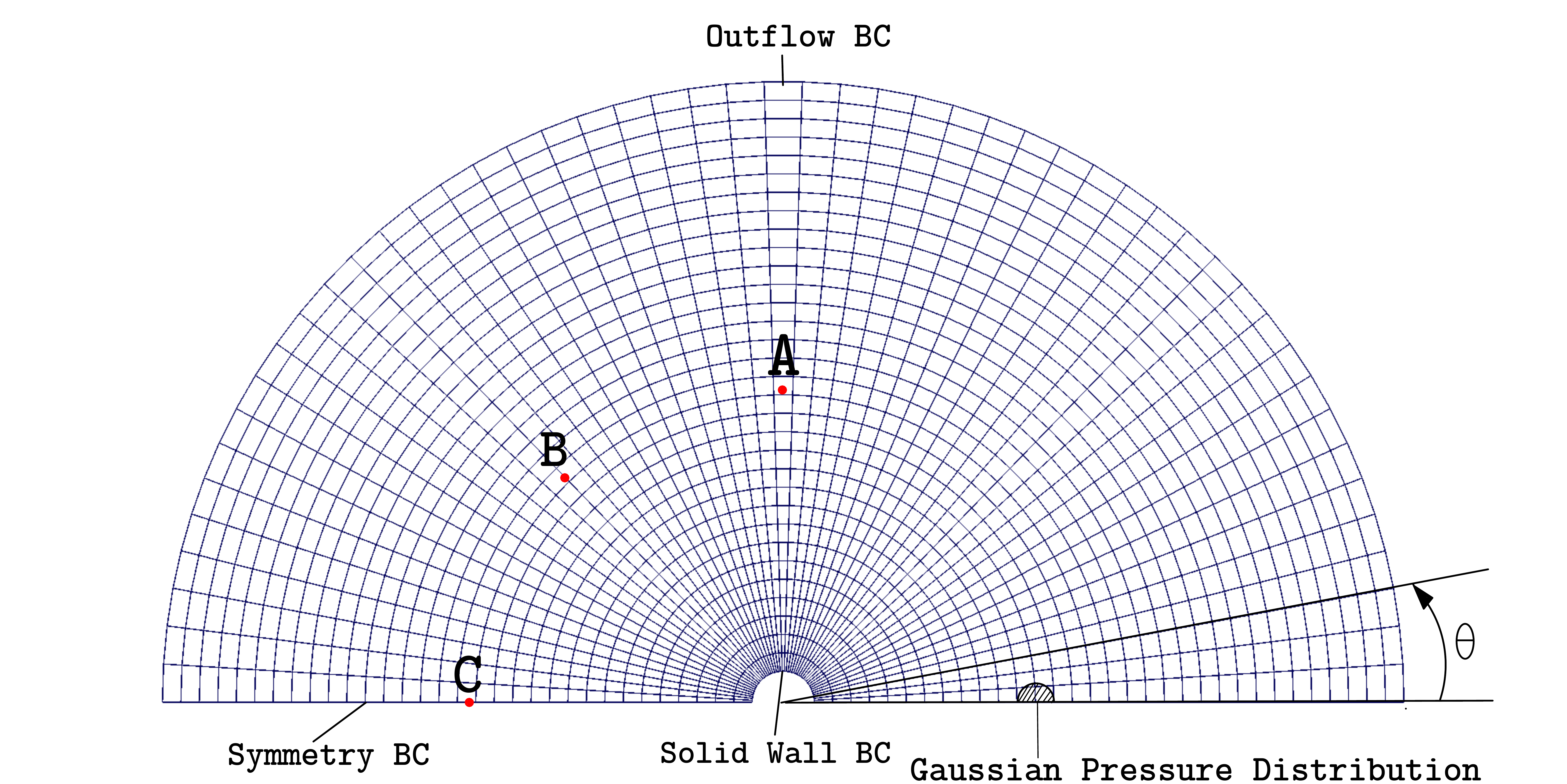}
	\end{center}
	\caption{\small{Schematic of geometry of the problem and computational grid}}
	\label{scheme}
\end{figure}

At $t>0$, an acoustic wave originating at $x=(4,0)$ propagates in all directions. 
This wave reflects from a
circular cylinder of radius $0.5$ units situated at $(x,y) = (0,0)$. 
Perturbation pressure at a point in the far-field 
is recorded at regular time intervals.
For this simulation, an axisymmetric grid consisting of  
$401$ cells in $\theta$ direction and $201$ cells in radial direction is used.
Only half of the domain is considered 
because of the geometrical symmetry of the problem.
A schematic of the geometry and the grid is shown in Fig.\ref{scheme}.
A coarser grid is selected in order to amplify the differences in the numerical solutions of different schemes.

\begin{figure}[H]
	\begin{center}
		\includegraphics [width=0.8\textwidth]{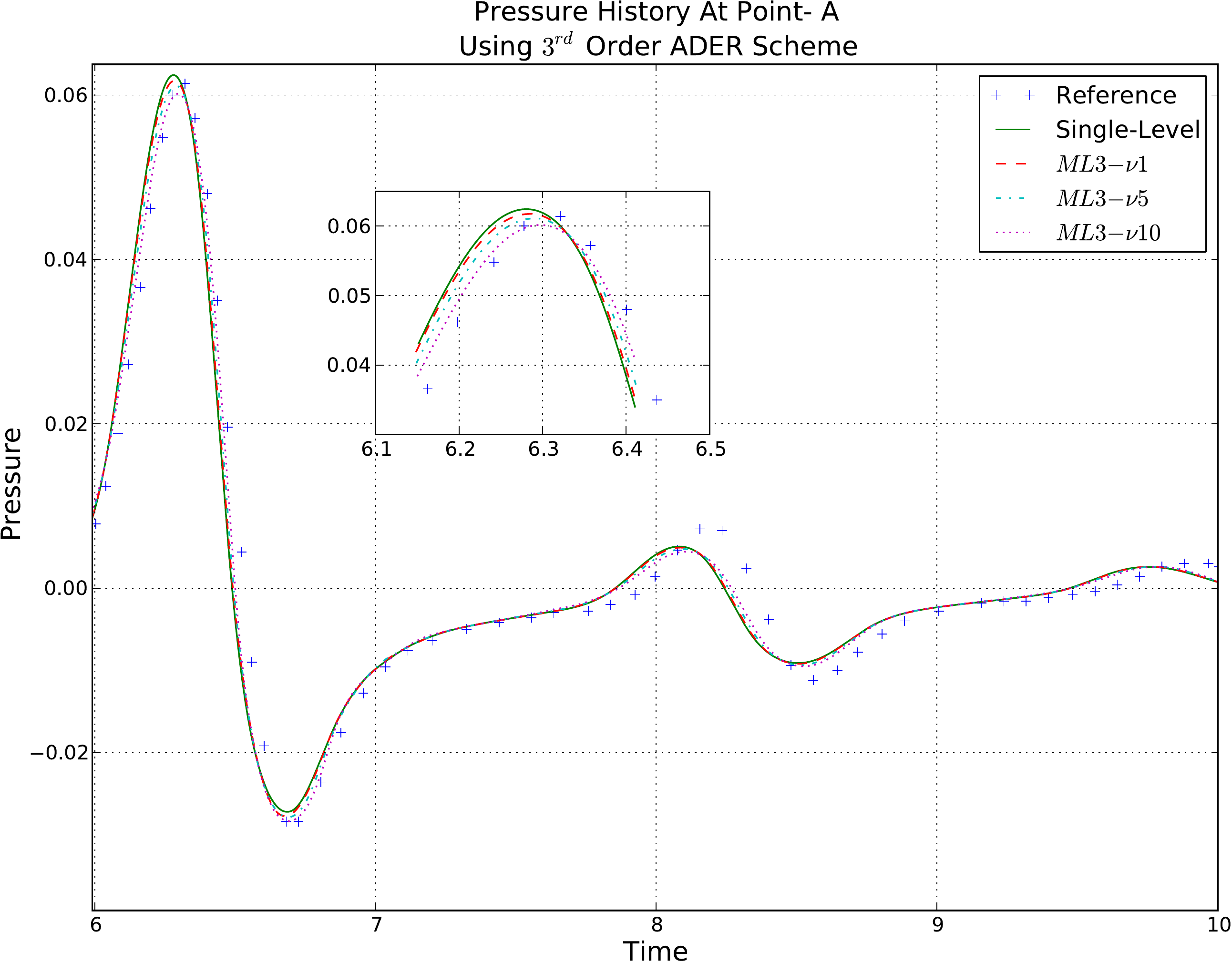}
	\end{center}
	\caption{\small{Comparison of the conventional and multilevel ADER methods with the reference solution \cite{CAA2}: Pressure history at a point in far-field due to acoustic scattering from a $2$D cylinder}}
	\label{history}
\end{figure}

\begin{figure}[H]
	\begin{minipage}[t]{0.5 \textwidth}
			\includegraphics [scale=0.41]{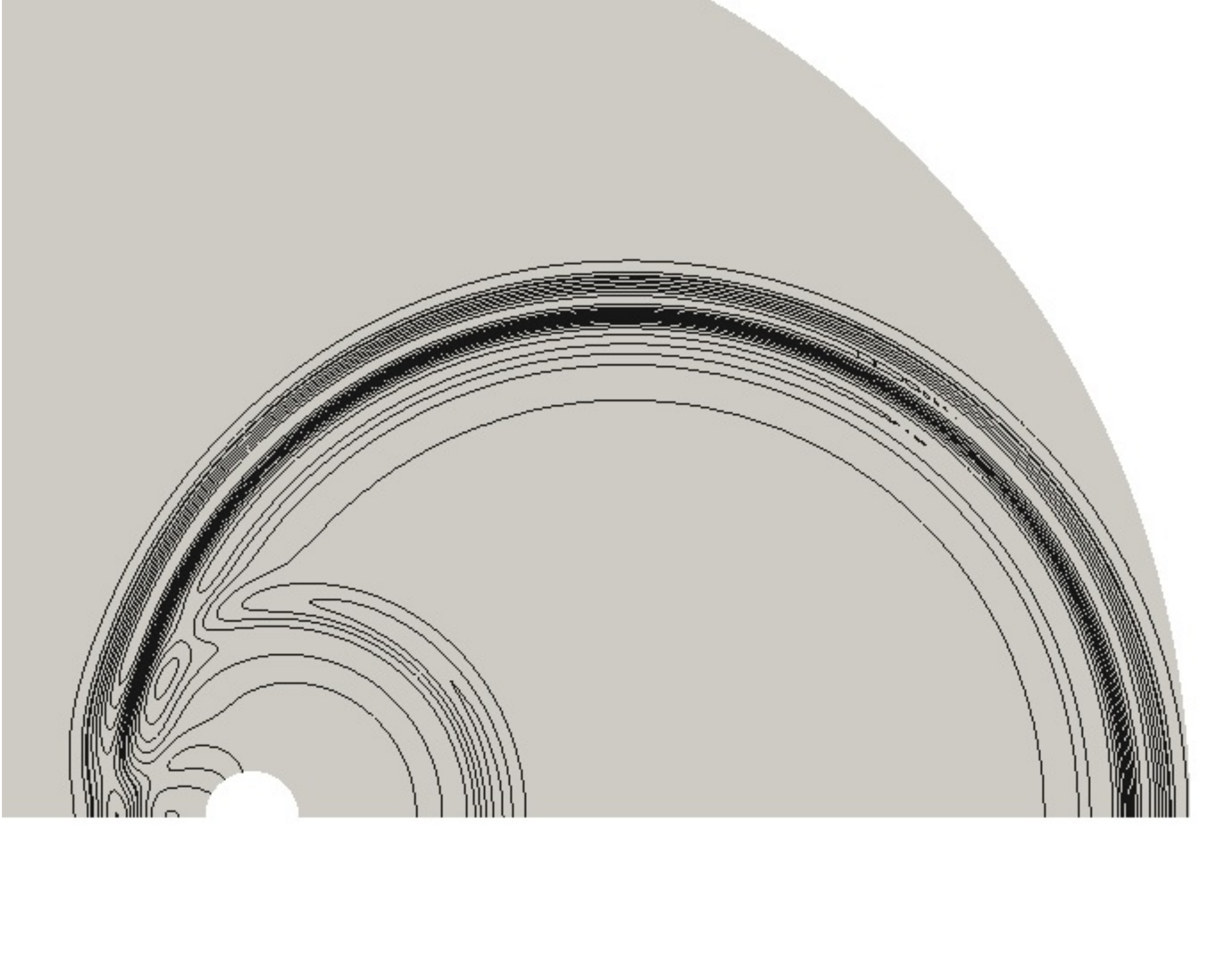}
	\end{minipage}%
	\begin{minipage}[t]{0.5 \textwidth}
			\includegraphics [scale=0.41]{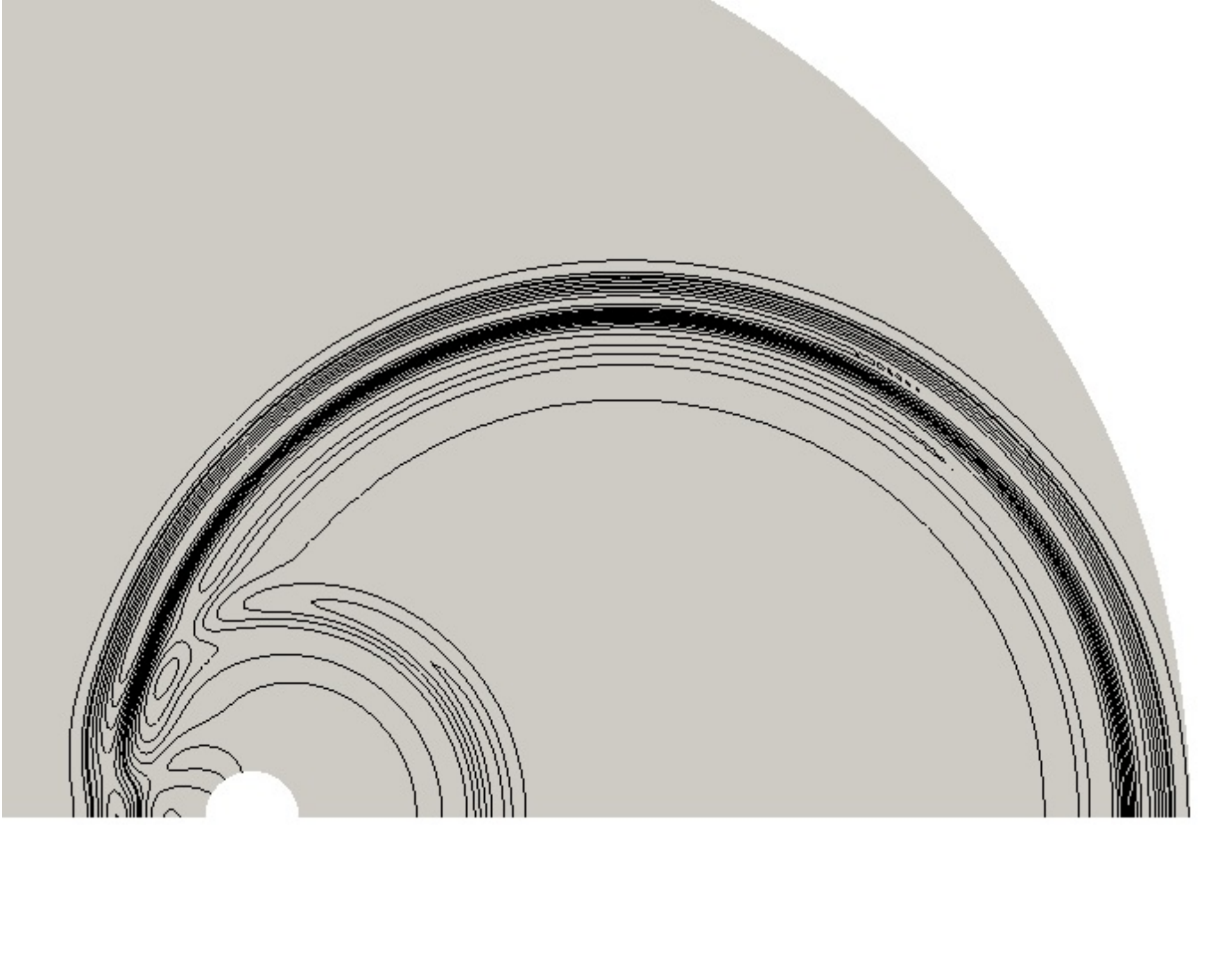}
	\end{minipage}%
	\caption{\small{Acoustic scattering from a $2$D circular cylinder: A. Pressure-contour plot with traditional ADER scheme B. Contour plot with ML-ADER method with $\nu=1$}}
	\label{Fig3}
\end{figure}

Fig.\ref{history} shows pressure history at a point $A (r=5, \theta = 90^0)$ in far-field. 
Contour plot of pressure perturbations are shown in Fig.\ref{Fig3}.
It can be seen that, the results obtained with ML-ADER method (and frozen $\tau$ variants) 
agree very well with the results from a conventional ADER method.

\section{Computational Efficiency}

Theoretically, the multilevel algorithm results in fraction of the 
computing cost as that of the single-level method of the same spatial order \cite{Chatterjee2015}. 
Table \ref{saving} shows computing time (in seconds) for simulating linear travelling waves using
a scalar linear advection equation with single as well as multi-level methods. 
ENO reconstruction procedure is used in all simulations.  
All simulations are performed on Intel $i5-2500$ CPU ($3.30 $GHz) running Linux.  
CPU time shown in the Table \ref{saving} is an average of computing time for three runs of each
simulation. 
A sufficiently small time-step is chosen for all simulations to reduce temporal errors. 
As number of cells on domain
are increased, the computing time required for both ADER and ML-ADER increases. However, the time 
required for ML-ADER is significantly less than time for conventional ADER method. 
It is also noted that, the ML-ADER method is as accurate as the conventional ADER method.
Table \ref{saving} shows the L$1$ errors corresponding to ADER and ML-ADER methods. 
The ML-ADER method results in significantly reduced computing cost 
as compared to traditional ADER method for the same numerical accuracy.

\begin{table}[h]
\begin{center}
\begin{tabular}{ | c | l | l | l | l | l | l | l |}
	
	\hline 
	\multicolumn{8}{|c|}{\bf Computational Performance} \\
	\hline 
	{Order /} & { Number of } &\multicolumn{2}{|c|}{CPU Time (S)} &\multicolumn{2}{|c|}{L$1$ Error} &\ \ \ \%  & Average \\
	\cline{3-4} \cline{5-6} 
	  {  Levels }& { FV cells} &  { ADER} & { ML-ADER} &  { ADER} & { ML-ADER} & Saving & Saving (\%) \\
	\hline 
	2       &50  & 0.35  & 0.20    & 1.85E-003	& 1.85E-003 & 43.40 &       \\
		&100 & 1.36  & 0.78    & 4.82E-004	& 4.82E-004 & 42.75 & 44.94 \\
		&200 & 6.73  & 3.45    & 1.27E-004	& 1.27E-004 & 48.66 &       \\ 
	\hline 

	3       &50  & 1.00  & 0.47    & 3.30E-005	& 3.30E-005 & 52.82 &       \\
		&100 & 4.52  & 1.94    & 4.13E-006	& 4.17E-006 & 57.08 & 56.18 \\
		&200 & 25.20 & 10.43   & 5.16E-007	& 5.54E-007 & 58.62 &       \\
	\hline 

	4       &50  & 2.59  & 0.99    & 4.85E-006	& 4.76E-006 & 61.73 &       \\
		&100 & 11.34 & 4.44    & 2.96E-007	& 2.94E-007 & 60.85 & 61.14 \\
		&200 & 60.53 & 23.69   & 2.62E-008	& 3.87E-008 & 60.86 &       \\
	\hline 
\end{tabular}
\caption{\label{saving}Saving in computing time with Multilevel ADER methods}
\end{center}
\end{table}

\begin{table}[h]
\begin{center}
\begin{tabular}{ | c | c | c | c | c | c | c |}
	
	\hline 
	\multicolumn{7}{|c|}{\bf Computational Performance: Benchmark Problems} \\
	\hline 
	{} & \multicolumn{3}{|c|}{\bf Benchmark Problem $1$} & \multicolumn{3}{|c|}{\bf Benchmark Problem $2$} \\
	{Numerical} & \multicolumn{3}{|c|}{Quasi $1$D, Fixed Stencil Reconstruction} & \multicolumn{3}{|c|}{$2$D, ENO Reconstruction} \\
	\cline{2-4} \cline{5-7} 
	{  Scheme }& {Computing} &  {Normalized} & {\%} & {Computing} &  {Normalized} & {\%} \\

	{ }& {Time (S)} &  {Time} & {Saving} & {Time (S)} &  {Time} & {Saving} \\
	\hline 
	SL-ADER O($3$)     & 37.74 & 1.0000 & 0.00  & 317322 & 1.0000 & 0.00 \\
	ML-ADER $\nu = 1$  & 23.61 & 0.6256 & 37.44 & 196175 & 0.6182 & 38.18 \\
	ML-ADER $\nu = 5$  & 14.74 & 0.3906 & 60.94 & 125668 & 0.3960 & 60.40\\
	ML-ADER $\nu = 10$ & 11.98 & 0.3174 & 68.26 & 104029 & 0.3278 & 67.22  \\

	\hline 
\end{tabular}
\caption{\label{savingBM}Saving in computing time with Multilevel ADER methods: Benchmark problems in CAA}
\end{center}
\end{table}

Table \ref{savingBM} shows computing time for conventional (SL) and multilevel (ML) ADER methods
for the Quasi-$1$D and $2$D benchmark problems in CAA. In this table, $\nu$ is
the number of frozen time-steps at the lowest ($1^{st}$) order of accuracy. 
Parameters such as the Courant number, number of cells and reconstruction procedure are 
kept identical for SL-ADER and ML-ADER in benchmark problems. 
It is seen that the multilevel method can result in significant saving in computing time.
Additional saving in computing time can be obtained with the Frozen-$\tau$ method (i.e. $\nu > 1$).

\section{Conclusion}
In this work, a multilevel framework for the versatile ADER method is presented. 
Theoretical and numerical analysis of the resulting ML method is carried out in frequency and
time domain. The ML-ADER method 
does not require restriction and prolongation operators and blends easily with existing 
finite volume framework.  
Theoretically, the ML method retains higher-order accuracy and does not introduce spatial
errors for time-accurate advection if the solution is sufficiently smooth.
Numerical grid converge tests using both monochromatic and broadband initial conditions
confirm that the ML method maintains higher-order accuracy.
It is seen that, for larger time-step, diffusion may get added in the higher wavenumber
components of the solution. However, solution to benchmark problems in CAA indicate that
ML-ADER method results in a higher-order accurate solution even at larger time-steps for 
practical problems.

The computing cost corresponding to a higher-order spatial reconstruction varies non-linearly
and thus plays an important role in overall cost of the numerical scheme \cite{Zingg2000}.
The ML-ADER method can achieve higher-order spatial accuracy at much lower computational cost.
The saving in overall computing cost increases with increasing computing load 
(such as reconstruction order or number of grid points). 
The computing cost can be further reduced by using the frozen-$\tau$ variant of the ML method.
Thus, the ML-ADER method can be effectively used for solving problems in CAA.
It may be possible to further increase computation efficiency by considering alternative cycling 
strategies as in conventional MG methods.

\bibspacing=\dimen 100
\bibliographystyle{unsrt}
\bibliography{Reference-ML_ADER_journal}

\end{document}